\documentclass[pra,twocolumn,superscriptaddress]{revtex4-2}

\usepackage{nicefrac,amsmath,amssymb,mathrsfs}
\usepackage{graphicx}
\usepackage{natbib}
\usepackage{bm}
\usepackage{xspace}
\usepackage{stmaryrd}
\usepackage{xifthen}

\begin{document}

\title{Tomographic reconstruction of quasistatic surface polariton fields}

\author{Raphael Hauer}
\affiliation{Graz Centre for Electron Microscopy, Steyrergasse 17, 8010 Graz, Austria}
\affiliation{Institute for Electron Microscopy and Nanoanalysis, Graz University of Technology,
Steyrergasse 17, 8010 Graz, Austria}

\author{Georg Haberfehlner}
\affiliation{Graz Centre for Electron Microscopy, Steyrergasse 17, 8010 Graz, Austria}
\affiliation{Institute for Electron Microscopy and Nanoanalysis, Graz University of Technology,
Steyrergasse 17, 8010 Graz, Austria}

\author{Gerald Kothleitner}
\affiliation{Graz Centre for Electron Microscopy, Steyrergasse 17, 8010 Graz, Austria}
\affiliation{Institute for Electron Microscopy and Nanoanalysis, Graz University of Technology,
Steyrergasse 17, 8010 Graz, Austria}

\author{Mathieu Kociak}
\affiliation{Universit\'e Paris-Saclay, CNRS, Laboratoire de Physique des Solides, 91405 Orsay. France}

\author{Ulrich Hohenester}
\affiliation{Institute of Physics, University of Graz, Universit\"atsplatz 5, 8010 Graz, Austria}

\date{\today}

\begin{abstract}
We theoretically investigate the tomographic reconstruction of the three-dimensional photonic environment of nanoparticles.  As input for our reconstruction we use electron energy loss spectroscopy (\textsc{eels}) maps for different rotation angles.  We perform the tomographic reconstruction of surface polariton fields for smooth and rough nanorods, and compare the reconstructed and simulated photonic local density of states, which are shown to be in very good agreement.  Using these results, we critically examine the potential of our tomography scheme, and discuss limitations and directions for future developments. 
\end{abstract}
\keywords{} 
\maketitle

\section{Introduction}

Nano optics deals with light confinement at the nanoscale~\cite{novotny:06,hohenester:20}.  This is achieved by binding light to surface resonances of nanoparticles, such as surface plasmon polaritons for metallic nanoparticles~\cite{maier:07} or surface phonon polaritons for dielectric nanoparticles~\cite{kliewer:74,caldwell:15}.  These resonances come along with strongly localized fields and allow squeezing light into extreme sub-wavelength volumes, which can be exploited for various applications~\cite{barbillon:19}.  

Because of the diffraction limit of light, the strongly localized fields cannot be directly imaged in optical microscopy.  In recent years, electron energy loss spectroscopy (\textsc{eels}) has become a highly successful technique for imaging electromagnetic fields at the nanoscale and with high energy resolution~\cite{nelayah:07,kociak:14,colliex:16,polman:19}.  In \textsc{eels} swift electrons pass by or through a nanoparticle and loose with a certain probability energy by exciting surface resonances.  By raster-scanning the electron beam over the specimen and measuring the number of electrons that have lost a certain amount of energy, one obtains information about the electromagnetic fields at the nanoscale~\cite{garcia:10,hohenester:20}.  However, the technique does not provide direct information about the three-dimensional fields but only about the averaged interaction along the entire electron trajectory.

\textsc{eels} tomography is a variant of electron tomography~\cite{midgley:09}, where the three-dimensional structure of a specimen is reconstructed from a collection of transmission electron micrographs for various tilt angles.  In \textsc{eels} the reconstruction is complicated by the fact that the loss does not occur at a specific position of the specimen, but is a highly nonlocal process~\cite{garcia:10}.  \textsc{eels} tomography of surface plasmons was first suggested independently in~\cite{hoerl.prl:13} and \cite{nicoletti:13}, where the latter paper demonstrated experimentally the reconstruction of localized surface plasmon modes for a silver nanocube.  While these seminal papers employed the quasistatic approximation~\cite{garcia:10,hohenester:20}, successive work showed how to extend the scheme to full retardation ~\cite{hoerl:15}, and demonstrated its applicability for single and coupled silver nanoparticles~\cite{hoerl:17,haberfehlner:17}.

In a recent paper~\cite{li:21}, we have brought \textsc{eels} tomography from the optical to the mid-infrared regime, and have demonstrated experimentally the reconstruction of localized surface phonon polaritons for a MgO nanocube.  Contrary to surface plasmon polaritons, the use of the quasistatic approximation is perfectly justified for surface phonon polaritons sustained by nanoparticles with dimensions of a few hundred nanometers.  This considerably simplifies the methodology for the tomographic reconstruction.  While going full circle from the quasistatic tomography of surface plasmon polaritons in our initial work~\cite{hoerl.prl:13} to quasistatic tomography of surface phonon polaritons~\cite{li:21}, we have gained quite some understanding of the critical elements in \textsc{eels} tomography, and our approach has matured considerably.  Time is ripe for a critical re-examination and re-interpretation of our tomography scheme.

In this paper we present a theoretical study of \textsc{eels} tomography for prototypical dielectric nanoparticles.  We submit a tilt series of simulated \textsc{eels} maps to our tomography scheme, in order to extract parameters characterizing the nanophotonic environment.  For this parametrized photonic environment we compute the photonic local density of states (\textsc{ldos}) \cite{garcia:08,novotny:06,hohenester:20}, which is compared with independent simulation results.  From this comparison we examine the strengths and weaknesses of our tomographic reconstruction scheme.  

The photonic \textsc{ldos} is a concept borrowed from solid state physics, and accounts for the number of photonic modes per unit frequency and volume.  In free space the photonic \textsc{ldos} is~\cite{novotny:06,hohenester:20} (we use SI units throughout)
\begin{equation}\label{eq:ldos}
  \rho_0(\omega)=\frac{\omega^2}{\pi^2c^3}\,,
\end{equation}
where $\omega$ is the angular frequency and $c$ the speed of light.  The photonic \textsc{ldos} governs the power dissipated by an oscillating dipole through
\begin{equation}
  P_0=\frac{\omega^2p^2}{12\varepsilon_0}\rho_0(\omega)\,,
\end{equation}
where $p$ is the oscillator's dipole moment and $\varepsilon_0$ the free-space permittivity.  Alternatively, we can relate via $P_0=\hbar\omega\,\gamma_0$ the power dissipation to the decay rate $\gamma_0$ of a quantum emitter.  The concept of the photonic \textsc{ldos} comes to full glory in nanophotonics, where the light-matter interaction becomes dramatically enhanced through surface excitations of nanoparticles, such as surface plasmon or phonon polaritons.  The enhancement of the photonic \textsc{ldos} $\rho(\omega)$ can be in the range of hundreds to thousands in comparison to its free-space value $\rho_0(\omega)$~\cite{schuller:10}.  Correspondingly, quantum emitters can transfer energy to the nanophotonic environment more efficiently, and their decay rate or power dissipation $P$ is increased by the \textsc{ldos} enhancement according to
\begin{equation}\label{eq:ldosenhace}
  P:P_0=\rho:\rho_0\,.
\end{equation}
Below we will compute the \textsc{ldos} enhancement $\rho:\rho_0$ using the photonic environment reconstructed from \textsc{eels} maps.  It is obvious that electrons and oscillating dipoles couple quite differently to the nanophotonic environment.  For this reason, the \textsc{ldos} reconstruction from \textsc{eels} data is quite delicate and provides a stringent testbed for our tomography approach.

We have organized our paper as follows.  In Sec.~\ref{sec:theory} we present the theory and methodology of our tomographic reconstruction.  We have tried to keep the presentation as compact and brief as possible, and refer to the literature for the detailed derivations whenever possible.  Some technical issues are transfered to an appendix.  In Sec.~\ref{sec:results} we present the tomography results for smooth and rough nanorods, and compare the reconstructed and the simulated photonic \textsc{ldos}.  Finally, in Sec.~\ref{sec:discussion} we put our tomography into a broader context, examine critically the strengths and weaknesses of our approach, and identify lines for future research.

\section{Theory}\label{sec:theory}

For MgO nanoparticles the surface phonon polariton energies $h\nu$ are of the order of 100 meV, corresponding to a free-space wavelength $\lambda=\nicefrac c\nu\sim 12\,\mu$m.  For nanoparticle dimensions of approximately hundred nanometers we can thus safely introduce the quasistatic approximation~\cite{hohenester:20}, where the electric field is expressed in terms of a quasistatic potential $V(\bm r)$ through $\bm E(\bm r)=-\nabla V(\bm r)$ and we keep the frequency dependence of the permittivity functions $\varepsilon(\omega)$.

\subsection{Green's functions}

\begin{figure}[t]
\includegraphics[width=\columnwidth]{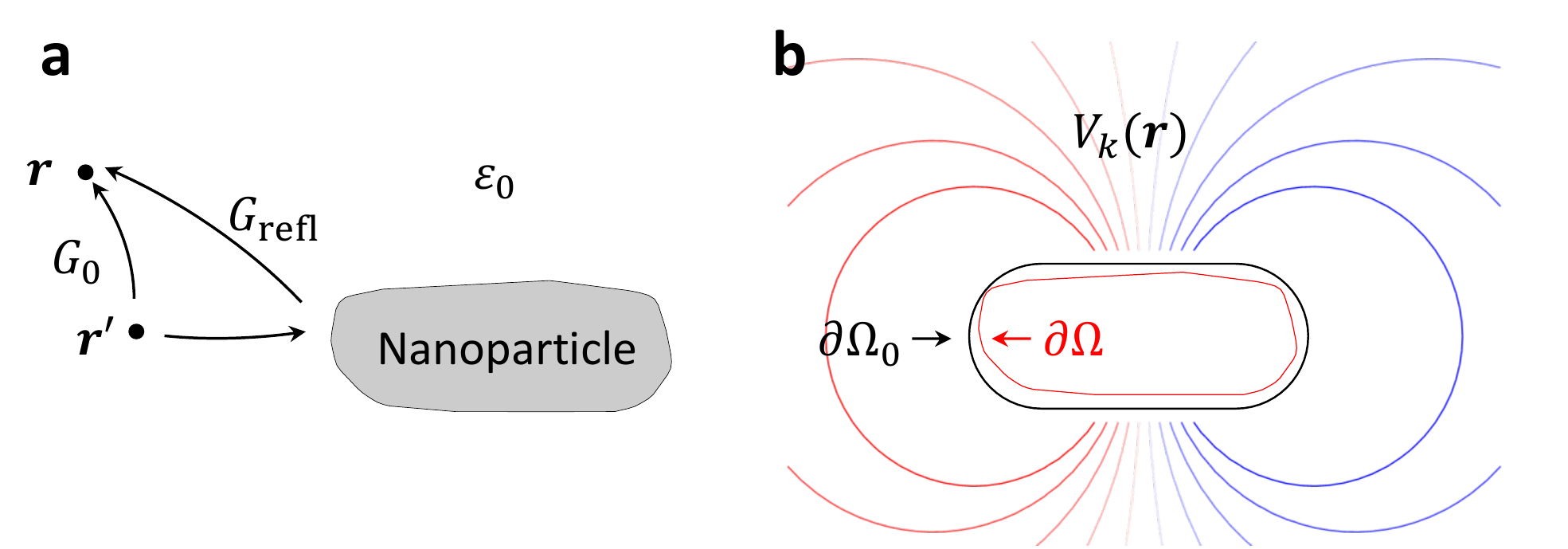}
\caption{(a) Schematics of Green's function.  In free space the Green's function $G_0(\bm r,\bm r')$ gives the potential at position $\bm r$ for a unit charge located at position $\bm r'$.  In presence of a nanoparticle one must additionally add a reflected Green's function that accounts for the nanoparticle response.  (b) The reflected Green's function can be expanded using a complete set of eigenpotentials $V_k(\bm r)$.  In our tomography scheme we can also start from the modes associated with a simpler reference boundary $\partial\Omega_0$ rather than the actual nanoparticle boundary $\partial\Omega$, and expand the reflected Green's function using the reference modes.  For details see text.}\label{fig:green}
\end{figure}

In the following we consider the problem depicted in Fig.~\ref{fig:green}(a), where a charge located at position $\bm r'$ interacts with a dielectric nanoparticle situated in a background medium with dielectric constant $\varepsilon_0$.  Green's functions provide an elegant and efficient method for solving such problems.  We first introduce the Green's function defined through~\cite{jackson:99,hohenester:20}
\begin{equation}\label{eq:green1}
  \nabla^2 G(\bm r,\bm r')=-\delta(\bm r-\bm r')\,,
\end{equation}
which gives the potential at position $\bm r$ for a unit charge located at position $\bm r'$.  In an unbounded medium the Green's function would be given by the usual expression
\begin{equation}\label{eq:green2}
  G_0(\bm r,\bm r')=\frac 1{4\pi|\bm r-\bm r'|}\,,
\end{equation}
and the potential associated with a charge distribution $\rho(\bm r)$ can be expressed as
\begin{equation}\label{eq:vinc}
  V_{\rm inc}(\bm r)=\int \frac{\rho(\bm r')}{4\pi\varepsilon_0|\bm r-\bm r'|}\,d^3r'\,.
\end{equation}
In presence of the nanoparticle this incoming potential will induce a reflected potential associated with the particle response.  To account for this, we split the total Green's function into two parts
\begin{equation}\label{eq:green3}
  G(\bm r,\bm r')=G_0(\bm r,\bm r')+G_{\rm refl}(\bm r,\bm r')\,,
\end{equation}
where the reflected part is a solution of Laplace's equation which is chosen such that Maxwell's boundary conditions are fulfilled at the nanoparticle boundary.  Suppose for a moment that the reflected Green's function is at hand.  It can then be shown that in \textsc{eels} the loss probability is related to the reflected Green's function via~\cite{garcia:10,hohenester:20}
\begin{equation}\label{eq:eels1}
  \Gamma(\bm R_0,\omega)=-\frac 1{\pi\hbar}\int \mbox{Im}\Big[\rho_{\rm el}^*(\bm r)
  G_{\rm refl}(\bm r,\bm r')\rho_{\rm el}(\bm r')\Big]\,d^3rd^3r'\,,
\end{equation}
where $\bm R_0=(x_0,y_0)$ is the impact parameter of the electron beam propagating along the $z$ direction (aloof geometry), $\hbar\omega$ is the loss energy and $\rho_{\rm el}(\bm r)$ the charge distribution of the swift electron.  The term in brackets of Eq.~\eqref{eq:eels1} accounts for a self-interaction process where the swift electron polarizes the nanoparticle, and the polarization acts back on the electron.  This nonlocal response is mediated by the reflected Green's function.  Similarly, the power dissipated by a dipole oscillating with frequency $\omega$ becomes~\cite{hohenester:20}
\begin{equation}\label{eq:ldos1}
  P=P_0-\frac\omega{2\varepsilon_0} \mbox{Im}\Big[
  (\bm p\cdot\nabla)(\bm p\cdot\nabla')G_{\rm refl}(\bm r,\bm r')\Big]_{\bm r=\bm r'=\bm r_0}\,,
\end{equation}
where $P_0$ is the free-space dissipation, $\bm p$ is the dipole moment, and $\bm r_0$ the position of the dipole.  The ratio $P:P_0$ gives the enhancement of the photonic \textsc{ldos}, see also Eq.~\eqref{eq:ldosenhace}.  The expressions given in Eqs.~\eqref{eq:eels1} and \eqref{eq:ldos1} are two examples for the enhancement of light-matter interactions in presence of nanoparticles, and show that the nanophotonic environment is fully characterized upon knowledge of the reflected Green's function.

\subsection{Eigenmode decomposition}

A powerful and convenient representation of the reflected Green's function is in terms of geometric eigenmodes $u_k(\bm s)$ and eigenvalues $\lambda_k$, where $\bm s$ is a position located on the boundary of the nanoparticle~\cite{ouyang:89,boudarham:12,hohenester:20}.  These eigenmodes form a complete set of basis functions.  To each eigenmode we can associate an eigenpotential
\begin{equation}\label{eq:eigpot}
  V_k(\bm r)=\oint_{\partial\Omega}\frac{u_k(\bm s')}{4\pi|\bm r-\bm s'|}\,dS'\,,
\end{equation}
which is a solution of Laplace's equation that fulfills Maxwell's boundary conditions at the nanoparticle boundary.  We can then decompose the reflected Green's function outside the nanoparticle in terms of these eigenpotentials via~\cite{boudarham:12,hohenester:20}
\begin{equation}\label{eq:grefl1}
  G_{\rm refl}(\bm r,\bm r')=-\sum_k V_k(\bm r)
  \left[\frac{\lambda_k+\frac 12}{\Lambda(\omega)+\lambda_k}\right]V_k(\bm r')\,,
\end{equation}
where $\Lambda(\omega)$ is an expression that solely depends on the permittivities of the nanoparticle and the embedding medium.  Inserting Eq.~\eqref{eq:grefl1} into the \textsc{eels} loss probability of Eq.~\eqref{eq:eels1} leads us to
\begin{equation}\label{eq:eels2}
  \Gamma(\bm R_0,\omega)=\frac 1{\pi\hbar\varepsilon_0}\sum_k
  L_k(\omega)\left|\int \rho_{\rm el}(\bm r)V_k(\bm r)\,d^3r\right|^2\,,
\end{equation}
with the lineshape function
\begin{equation}
  L_k(\omega)=\mbox{Im}\left[\frac{\lambda_k+\frac 12}{\Lambda(\omega)+\lambda_k}\right]\,.
\end{equation}
Eq.~\eqref{eq:eels2} is a particularly useful decomposition of the loss probability in terms of surface phonon polariton eigenmodes.  Each eigenmode contributes with the lineshape function $L_k(\omega)$ and the oscillator strength given by the square modulus term, which is governed by the interaction energy between the charge distribution of the swift electron and the eigenpotential $V_k(\bm r)$.  Similarly, the power dissipated by an oscillating dipole of Eq.~\eqref{eq:ldos1} can be decomposed into eigenmodes via
\begin{equation}\label{eq:ldos2}
  P=P_0+\frac{\omega}{2\varepsilon_0}\sum_k L_k(\omega)\big|\bm p\cdot\nabla V_k(\bm r)\big|^2_{\bm r=\bm r_0}\,,
\end{equation}
with a corresponding interpretation in terms of lineshape functions and oscillator strengths.  From the dissipated power one can obtain the photonic \textsc{ldos} using Eqs.~\eqref{eq:ldos} and \eqref{eq:ldosenhace}, where one often additionally averages over all dipole orientations to account for the random orientation of quantum emitters in typical experiments~\cite{novotny:06}.

\subsection{Tomographic reconstruction of eigenmodes}

It is apparent from Eqs.~\eqref{eq:eels2} and \eqref{eq:ldos2} that we can compute the \textsc{eels} loss probability $\Gamma(\bm R_0,\omega)$ and the \textsc{ldos} enhancement $P:P_0$, or any other related response function, once the geometric eigenmodes $u_k(\bm s)$ and the lineshape function $L_k(\omega)$ are at hand.  Expressed differently, the nanophotonic environment is fully characterized upon knowledge of $u_k(\bm s)$ and $L_k(\omega)$.  We can now formulate the goal of our tomography approach.  Suppose that we are in possession of the \textsc{eels} loss probabilities $\Gamma(\bm R_0,\omega)$, ideally for various impact parameters and electron propagation directions, but don't know the eigenmodes $u_k(\bm s)$ and lineshape functions $L_k(\omega)$: can we obtain through solution of an inverse problem a viable approximation for $u_k(\bm s)$ and $L_k(\omega)$?  And if yes, how?

\begin{figure*}[t]
\centerline{\includegraphics[width=1.6\columnwidth]{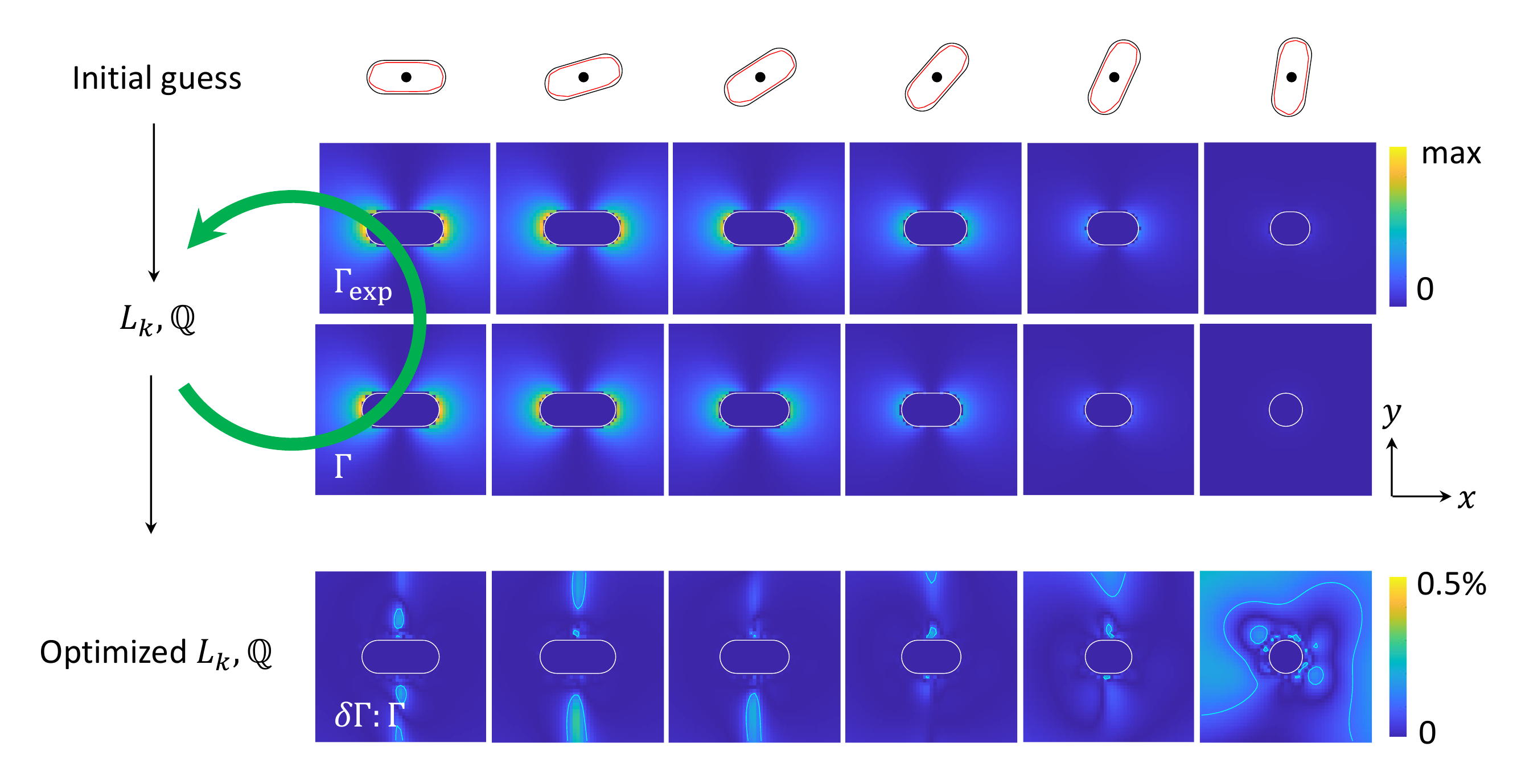}}
\caption{Schematics of tomographic reconstruction for a rough nanorod.  The reference boundary is formed by a smooth rod, see panels on top of the figure.  The experimental \textsc{eels} maps $\Gamma_{\rm exp}$ are obtained for a specific loss energy and for various rotation angles, we only keep aloof electron trajectories that do not penetrate the smooth rod.  We start with some initial guess for the optimization parameters $L_k$, $\mathbb{Q}$ and compute the reprojected \textsc{eels} maps $\Gamma$ using Eq.~\eqref{eq:eels3}.  These parameters are optimized until a local minimum is reached by the optimization algorithm.  In the lowest row we show the relative error $|\Gamma_{\rm exp}-\Gamma|:\Gamma_{\rm exp}$ between the experimental and optimized maps.  The solid lines indicate the contours for an error of~0.1\%.  Once the parameters $L_k$, $\mathbb{Q}$ are at hand, we can compute other quantities such as the photonics \textsc{ldos}.}\label{fig:cost}
\end{figure*}

\subsubsection{Optimization for modes on nanoparticle boundary}

Consider first the situation that the nanoparticle boundary is known and that we are seeking for the linshape functions and eigenmodes $L_k$, $u_k(\bm s)$.  This corresponds to the situation previously investigated in~\cite{li:21}.  Let $u_\ell^0(\bm s)$ be a complete set of basis functions on the boundary.  We shall refer to these modes as \textit{reference modes}.  As shown in Appendix~\ref{sec:ortho}, the eigenpotentials of Eq.~\eqref{eq:eigpot} can be expanded in terms of these modes via
\begin{equation}\label{eq:eigpotref1}
  V_k(\bm r)=\sum_\ell\mathbb{Q}_{k\ell}\oint_{\partial\Omega}
  \frac{ u^0_\ell(\bm s')}{4\pi|\bm r-\bm s'|}\,dS'\,,
\end{equation}
with $\mathbb{Q}$ being an orthogonal matrix.  We can now formulate the tomographic reconstruction scheme for a given set of experimental \textsc{eels} maps.

\begin{enumerate}

\item Find some reference modes $u_\ell^0(\bm s)$ whose gross features are expected to be similar to those of the true eigenmodes $u_k(\bm s)$.  This point is irrelevant for a complete basis, but becomes crucial for actual reconstructions where the basis has to be truncated.

\item Start with some initial guess for the lineshape function $L_k$ and orthogonal matrix $\mathbb{Q}$, and compute the reprojected maps via Eq.~\eqref{eq:eels2}.  Use an optimization routine for $L_k$, $\mathbb{Q}$ to obtain the best possible agreement between experiment and reprojection.  Note that in principle $L_k(\omega)$ depends on frequency, but for a fixed loss energy the lineshape functions can be treated as mere numbers.

\item Use the optimized parameters to compute other quantities, such as the photonic \textsc{ldos}.
\end{enumerate}

\subsubsection{Optimization for modes on reference boundary}

The above scheme can be also generalized to cases where the true nanoparticle boundary $\partial\Omega$ is not known or is too complicated to be used in actual reconstructions.  We start by introducing a reference boundary $\partial\Omega_0$ that fully encapsulates the nanoparticle, see also Fig.~\ref{fig:green}(b). In our modified approach we are not aiming for a reconstruction of the eigenmodes $u_k(\bm s)$ themselves, but of the eigenpotentials of Eq.~\eqref{eq:eigpot} outside of the reference boundary.  There they can be expressed as generic solutions of Laplace's equation~\cite{jackson:99}
\begin{equation}\label{eq:eigpotref2}
  V_k(\bm r)=\oint_{\partial\Omega_0}\frac{\sigma_k(\bm s')}{4\pi|\bm r-\bm s'|}\,dS'\,,
\end{equation}
where $\sigma_k(\bm s)$ specifies the normal derivative of the potential on $\partial\Omega_0$ (von Neumann boundary condition).  We can now use a complete set of basis functions $u_\ell^0(\bm s)$ on $\partial\Omega_0$ for the expansion of $\sigma_k(\bm s)$ to arrive at
\begin{equation}\label{eq:eigpotref3}
  V_k(\bm r)=\sum_\ell\mathbb{Q}_{k\ell}\oint_{\partial\Omega_0}
  \frac{ u^0_\ell(\bm s')}{4\pi|\bm r-\bm s'|}\,dS'\,,
\end{equation}
where $\mathbb{Q}$ is a non-orthogonal matrix formed by the expansion coefficients.  The tomographic reconstruction can now be performed in complete analogy to the scheme presented above, with the only exception that $\mathbb{Q}$ has to be replaced by a non-orthogonal matrix.

\begin{figure*}[t]
\centerline{\includegraphics[width=2\columnwidth]{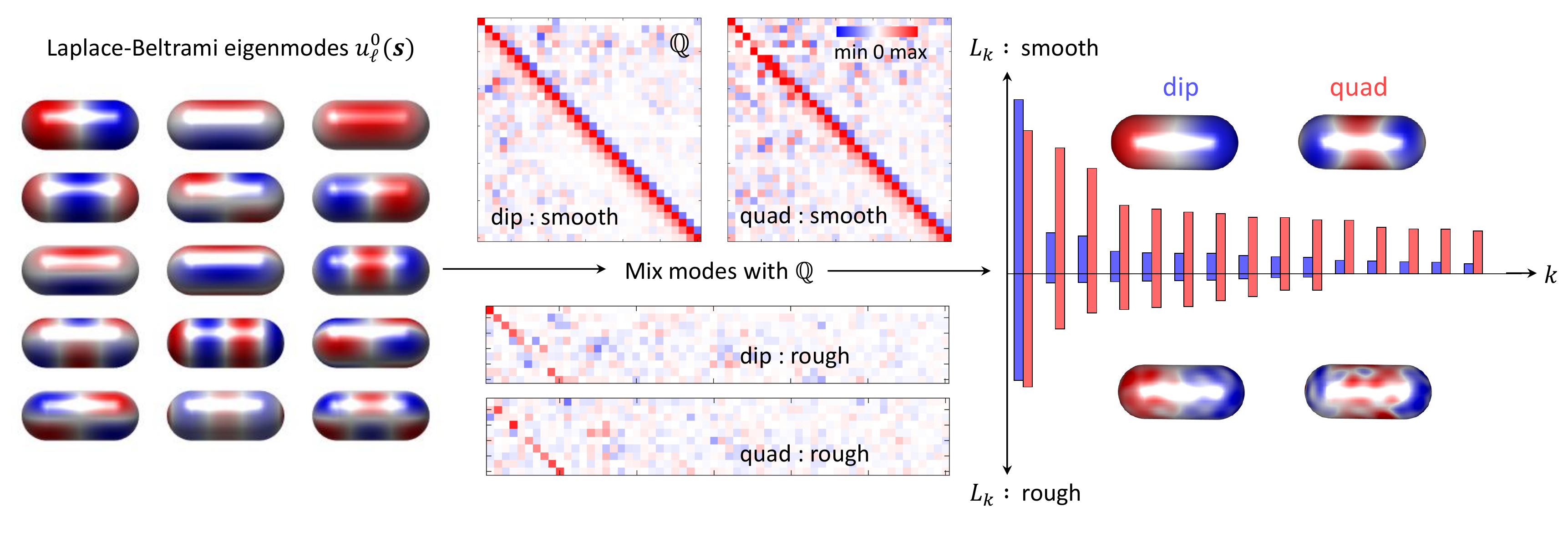}}
\caption{Reference and reconstructed modes.  In our tomographic reconstruction we use as reference modes $u_\ell^0(\bm s)$  the eigenmodes of the Laplace-Beltrami operator.  Using the optimized parameters $L_k$, $\mathbb{Q}$ we mix the modes to obtain the reconstructed modes shown on the right hand side for the dipole and quadrupole resonances.  For the smooth rod, $\mathbb{Q}$ is an orthogonal matrix of size $n\times n$, where $n$ is the truncation number of the basis.  For the rough rod, $\mathbb{Q}$ is a full matrix of size $m\times n$, where $m$ is the number of eigenpotentials to be reconstructed.  From the knowledge of $\mathbb{Q}$ we can compute the geometric eigenpotentials $V_k(\bm r)$ outside the reference boundary.  The bar plot on the right hand side reports the reconstructed lineshape parameters for the dipole (blue) and quadrupole (red) resonances, the modes are sorted in decreasing order of $L_k$ and the largest contributions are due to the modes shown in the insets.}\label{fig:modes}
\end{figure*}

\subsubsection{Optimization loop}

In the following we discuss the optimization procedure in slightly more detail, see also Figs.~\ref{fig:cost} and \ref{fig:modes}.  We provide a unified description for the optimizations using modes defined on either the nanoparticle or reference boundary.  In the first case, $\mathbb{Q}$ is an orthogonal matrix.  In our computational approach we have to truncate the basis and keep only $n$ representative modes, where $n$ is of the order of several tens to hundreds.  Correspondingly, $\mathbb{Q}_{n\times n}$ is a matrix of size $n\times n$, see also Appendix~\ref{sec:ortho} for the parametrization of this matrix.  In the case of a reference boundary, $\mathbb{Q}$ is a full matrix.  In principle we can now use different truncation numbers $m$ and $n$ for the reconstructed eigenpotentials and basis functions, respectively, and $\mathbb{Q}_{m\times n}$ becomes a matrix of size $m\times n$.  In most cases it is sufficient to consider around ten eigenpotentials, whereas the truncation number for the basis should be chosen considerably larger.  Let 
\begin{equation}
  x_i=\big\{\bm R_0^{(i)},\theta^{(i)}\big\}
\end{equation}
be a set of impact parameters and tilt angles for a fixed loss energy, and $\Gamma_{\rm exp}(x_i)$ the corresponding experimental \textsc{eels} maps.  We only consider aloof electron trajectories that do not penetrate the nanoparticle.  The interaction energy between the swift electron and a reference mode $u_\ell^0(\bm s)$ is
\begin{equation}
  \mathscr{V}_\ell(x_i)=\int\rho_{\rm el}(\bm r)V_\ell^0(\bm r)\,d^3r=
   \oint_{\partial\Omega_0}V_{\rm el}(\bm s)u_\ell^0(\bm s)\,dS\,,
\end{equation}
where $V_{\rm el}(\bm r)$ is the potential associated with the charge distribution $\rho_{\rm el}(\bm r)$.  When the nanoparticle boundary is known, the reference boundary in the above boundary integral is identical to $\partial\Omega$.  The loss probability of Eq.~\eqref{eq:eels2} can then be written in the compact form
\begin{equation}\label{eq:eels3}
  \Gamma(x_i;L_k,\mathbb{Q})=\frac 1{\pi\hbar\varepsilon_0}\sum_k\sum_{\ell,\ell'} 
  \big(\mathbb{Q}_{k\ell}\mathscr{V}_\ell(x_i)\big)L_k
  \big(\mathbb{Q}_{k\ell'}\mathscr{V}_{\ell'}(x_i)\big)\,.
\end{equation}
We can now define a cost function
\begin{equation}\label{eq:cost}
  J(L_k,\mathbb{Q})=\frac 12\sum_i\Big|\Gamma_{\rm exp}(x_i)-\Gamma(x_i;L_k,\mathbb{Q})\Big|^2
  \longrightarrow\mbox{min}\,,
\end{equation}
that gives the ``distance'' between the experimental and reprojected \textsc{eels} maps.  This cost function is  submitted to an optimization routine, such as a conjugate-gradient or quasi-Newton one~\cite{press:02}, which provides us with the optimized expressions for $L_k$, $\mathbb{Q}$.  Some details about the parametrization of the orthogonal matrix, as well as the computation of the derivative of the cost function with respect to the optimization parameters are given in Appendix~\ref{sec:ortho}.

\section{Results}\label{sec:results}

In~\cite{li:21} we have applied our tomography scheme to experimental \textsc{eels} maps for a MgO nanocube.  In this work we proceed differently and investigate the working principle of our tomography scheme using simulated data only.

\begin{enumerate}

\item We first compute for each loss energy \textsc{eels} maps for a series of rotation angles, see also Fig.~\ref{fig:cost}.  To be consistent with our previous notation, we denote these simulated \textsc{eels} maps as $\Gamma_{\rm exp}$, and will refer to them as \textit{experimental \textsc{eels} maps}.

\item These maps are submitted to our tomography scheme based on Eq.~\eqref{eq:cost}, in order to obtain the optimized parameters $L_k$, $\mathbb{Q}$ that specify the nanophotonic environment.

\item Using Eq.~\eqref{eq:ldos2} together with the optimized parameters, we compute the photonic \textsc{ldos}, and will refer to it as the \textit{reconstructed photonic} \textsc{ldos}.

\item Using Eq.~\eqref{eq:ldos1} we compute the photonic \textsc{ldos} directly, with a simulation approach to be discussed below, and will refer to it as the \textit{simulated photonic} \textsc{ldos}.  
\end{enumerate}

\begin{figure}[b]
\centerline{\includegraphics[width=\columnwidth]{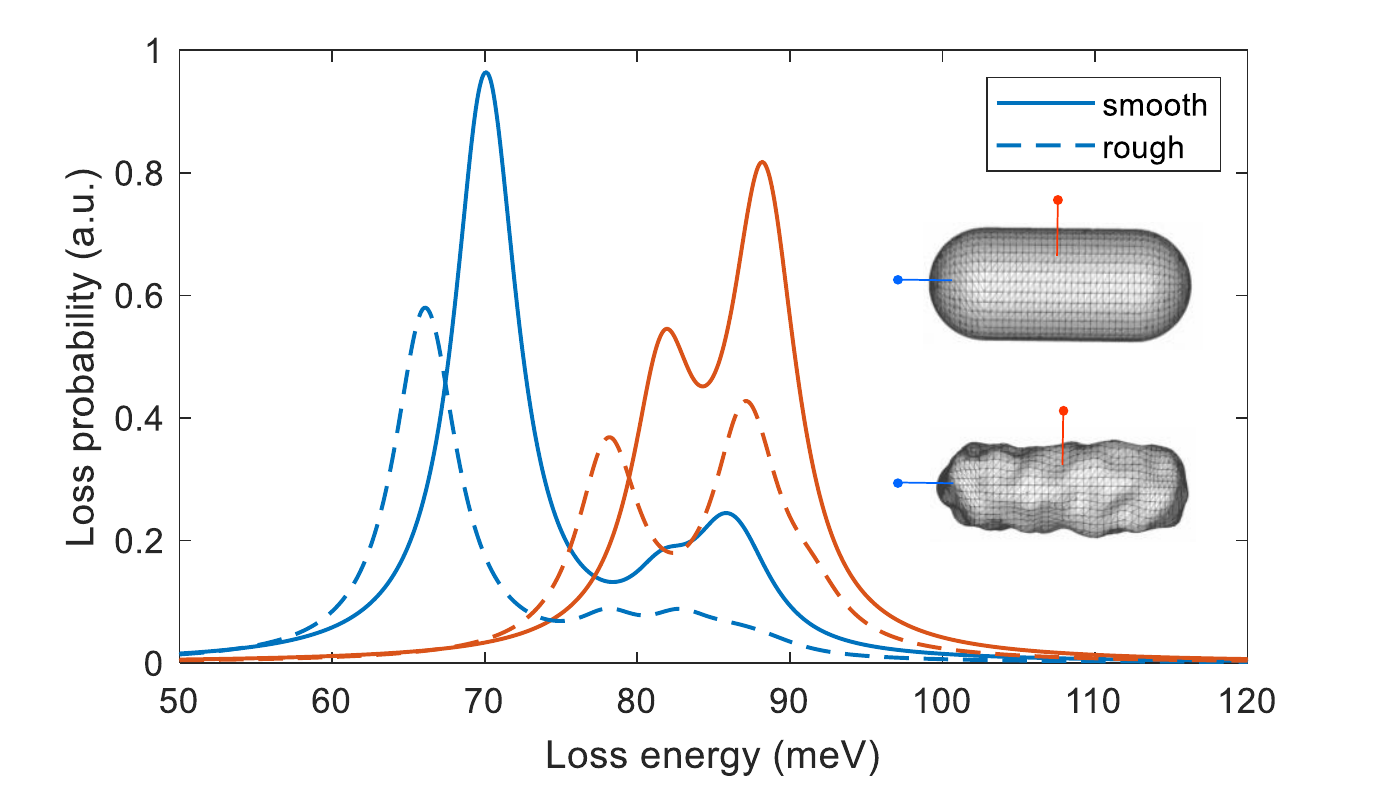}}
\caption{Loss spectra for smooth and rough nanorod, and for impact parameters located on the long (blue) and short (red) rod axis, see inset.  We consider aloof electron trajectories with a propagation direction out of the image plane.  One observes a dipole resonance around 70 meV, a quadrupole resonance around 80 meV, and a peak attributed to a multitude of modes around 90 meV.}\label{fig:spectra}
\end{figure}

\noindent For ideal reconstruction the simulated and reconstructed \textsc{ldos} maps should be identical.  Any deviation between the two maps can thus be attributed to deficiencies of our approach, caused for instance by the truncation of the reference basis $u_\ell^0(\bm s)$ or a trapping of the optimization algorithm in a local minimum.

We apply our tomography scheme to prototypical systems of a smooth and rough nanorod with a diameter to length ratio of approximately $1:2.5$, see also Fig.~\ref{fig:spectra} and \cite{lourenco:17} for a detailed discussion of the rod modes.  The rough rod has been by generated by adding stochastic height variations to the smooth surface of an ideal nanoparticle, following the prescription given in~\cite{truegler.prb:11}.  We shall not be concerned whether such nanoparticles can indeed be fabricated with the material system under investigation.  As we are working within the quasistatic regime, the actual size of the nanorods is irrelevant and the results can be easily scaled to any size.

\subsection{Computational details}

All our simulations are performed with the quasistatic classes of the \textsc{nanobem} toolbox~\cite{hohenester.cpc:22}, which is based on a Galerkin scheme with linear shape elements.  See e.g.~\cite{hohenester:20} for a detailed discussion.  The parametrization of the MgO dielectric function is the same as in~\cite{hohenester.prb:18,li:21}.  The nanorod boundaries are discretized using more than 3000 boundary elements of triangular shape.  We checked that for such fine discretizations we obtained converged results.  As for the \textsc{eels} simulations, we consider the limit of large electron velocities $v$ where the potential for a swift electron with impact parameter $\bm R_0$ takes the form
\begin{equation}
  V_{\rm el}(\bm r;\bm R_0)=-\frac{e}{2\pi v}\ln\big|\bm R-\bm R_0\big|\,,
\end{equation}
with $e$ being the elementary charge and $\bm R=(x,y)$.  We have previously shown in Fig.~S9 of \cite{li:21} that this simplified expression gives almost the same results as simulations based on the full Maxwell's equations.

As for the reference modes $u_\ell^0(\bm s)$, we did not choose the usual geometric eigenmodes~\cite{boudarham:12,hohenester:20} for two reasons.  First, in order to demonstrate that our approach indeed works for any meaningful set of basis functions.  Second, we observed that the geometric eigenmodes computed with the \textsc{nanobem} toolbox are often strongly localized around sharp corners or edges, such that a large number of such modes would be needed for a useful expansion.  In this work, we choose for $u_\ell^0(\bm s)$ the eigenmodes of the Laplace-Beltrami operator, which is a generalization of the Laplace operator for curved boundaries and is known to provide extremely smooth basis functions~\cite{bobenko:07}.  The modes were additionally orthogonalized using Eq.~\eqref{eq:orthoref}.

In our optimization approach we truncate the Laplace-Beltrami basis using the $n$ modes of highest eigenvalue, where a value of $n\approx 100$ turned out to be a good compromise between reasonably fast optimizations and sufficiently accurate results.  The optimization was performed with the builtin \textsc{matlab} function \texttt{fminunc} using a quasi-Newton algorithm together with a relatively small function and optimality tolerance of $10^{-8}$.  In all our simulation we typically needed about 2000 iterations to reach a local minimum.

\subsection{Smooth rod}

We start by discussing the smooth rod shown in Fig.~\ref{fig:spectra}.  The loss spectra exhibit three peaks, which can be attributed to a dipolar mode (70 meV), a quadrupolar mode (80 meV), and a peak that is composed of a multitude of modes (88 meV).  For the smooth rod, the reference boundary $\partial\Omega_0$ is identical to the true nanoparticle boundary $\partial\Omega$.  Note that the Laplace-Beltrami eigenmodes provide a (truncated) basis that does not coincide with the true geometric eigenmodes.  Simulated and reprojected maps originating from our optimization algorithm are typically extremely similar, see lowest row in Fig.~\ref{fig:cost} for the more difficult case of the rough nanorod.

\begin{figure}[t]
\centerline{\includegraphics[width=\columnwidth]{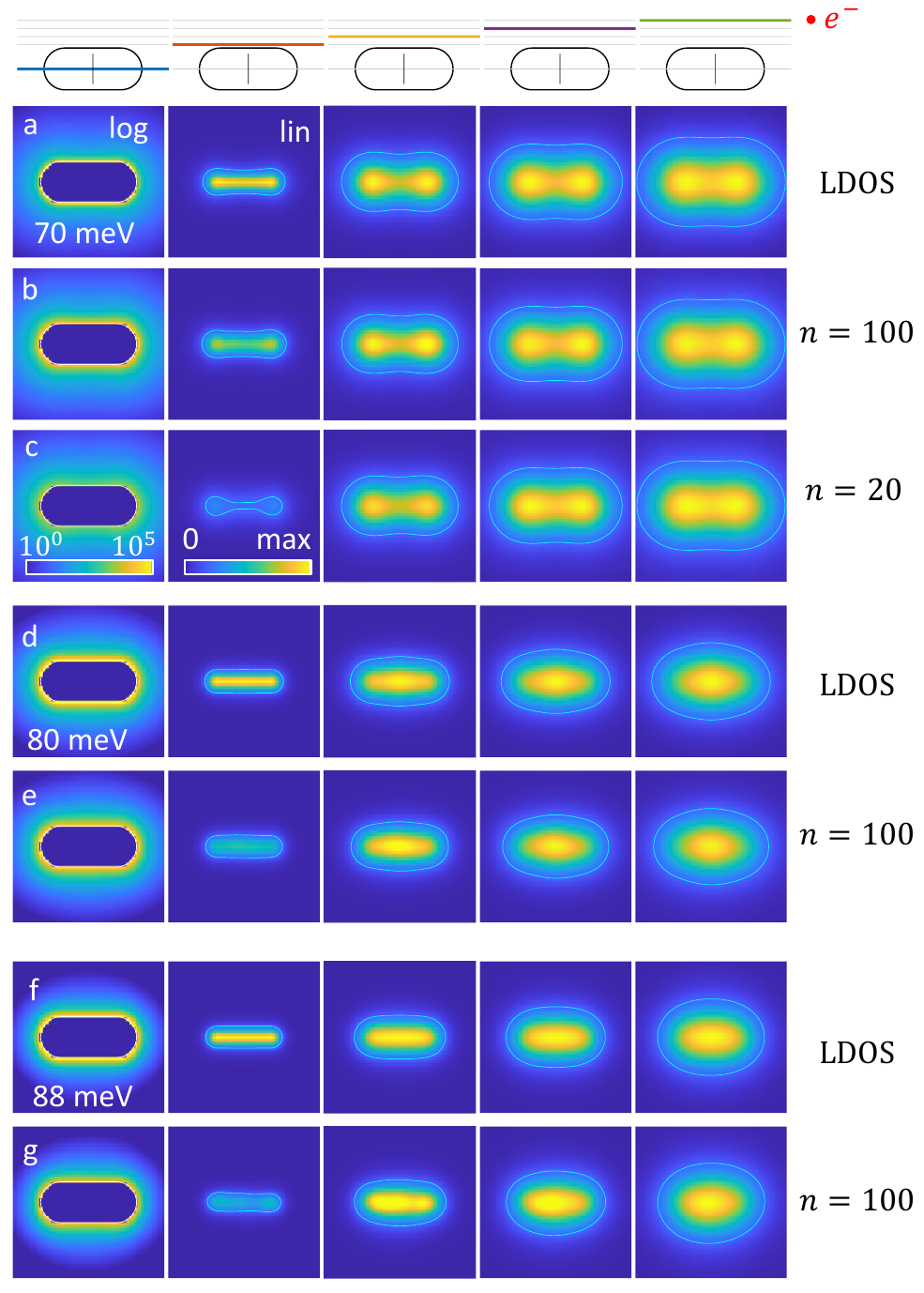}}
\caption{Simulated and reconstructed \textsc{ldos} maps for the different loss energies reported in the panels and in the different planes indicated on top of the figure.  The electron propagation direction is out of the image plane and the lines at the rod centers indicate the tilt axis.  (a) Simulated \textsc{ldos} maps for dipole mode, (b,c) reconstructed \textsc{ldos} maps for different numbers $n$ of Laplace-Beltrami eigenmodes.  Same for (d,e) quadrupole resonance and (f,g) multitude of modes.  The \textsc{ldos} maps in the first column are displayed for a logarithmic color scale, in the other columns we use a linear color scale.  All maps are scaled to the maxima of the simulated maps.  The solid lines report the contours for 20\% of the maximum of the simulated \textsc{ldos} in the respective planes.}\label{fig:ldos1}
\end{figure}

\begin{figure}[t]
\centerline{\includegraphics[width=\columnwidth]{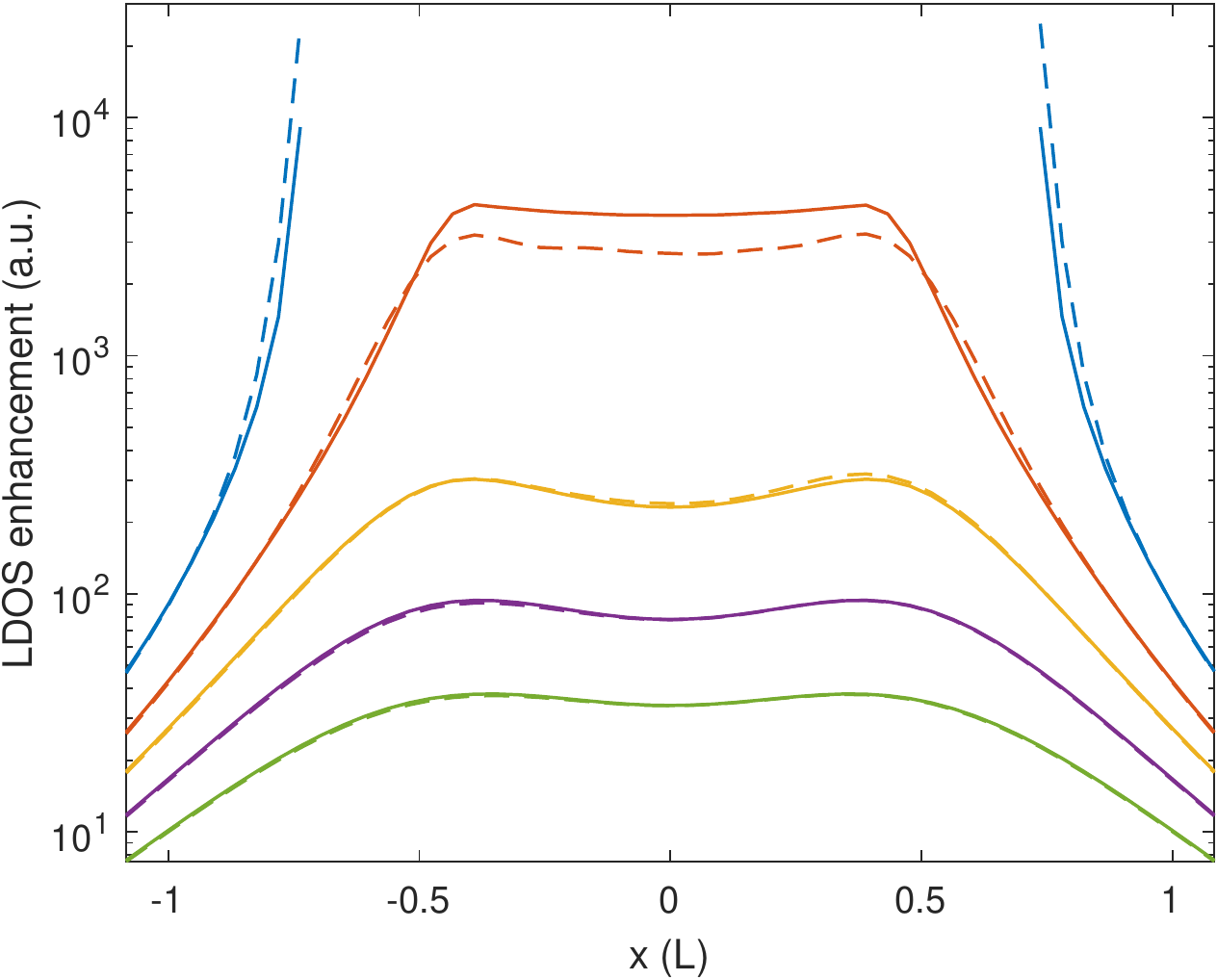}}
\caption{Cuts through the \textsc{ldos} maps shown in Figs.~\ref{fig:ldos1}(a,b) along the long rod axis at $y=0$.  The solid lines report simulation results, the dashed lines show the reconstructed results.  The \textsc{ldos} enhancements are given in arbitrary units, with a constant prefactor for the reconstructed \textsc{ldos} maps.  For a discussion see text.  Larger \textsc{ldos} enhancements correspond to positions closer to the nanorod, the colors are in agreement with those of the planes shown on top of Fig.~\ref{fig:ldos1}.  Distances are given in units of the rod length $L$.}\label{fig:ldos2}
\end{figure}

Figure~\ref{fig:ldos1} shows the simulated and reconstructed \textsc{ldos} maps in the symmetry plane (left column) and in planes away from the rod (other columns), and for the loss energies reported in the figure.  We first consider the dipole mode shown in panel~(a).  The \textsc{ldos} can be interpreted for an oscillating dipole as the enhancement of the dissipated power, see Eq.~\eqref{eq:ldos1}, throughout we average over all possible dipole orientations.  Close to the rod, an oscillating dipole couples with comparable strength to all surface phonon polariton modes.  This can be seen both in the symmetry plane of the rod (first column, logarithmic color scale), as well as in the plane closest to the rod (second column, linear color scale), where the photonic \textsc{ldos} is large and unstructured close to the rod boundary.  When moving away from the rod (other columns from left to right), the coupling strength between the oscillating dipole and the rod resonance modes have different distance dependencies, which are governed by the oscillator strengths given in Eq.~\eqref{eq:ldos2}.  For the chosen loss energy the dipolar rod mode becomes strongest at larger distances, as can be inferred from the two lobes in the \textsc{ldos} maps located at the rod caps.

\begin{figure*}[t]
\centerline{\includegraphics[width=2\columnwidth]{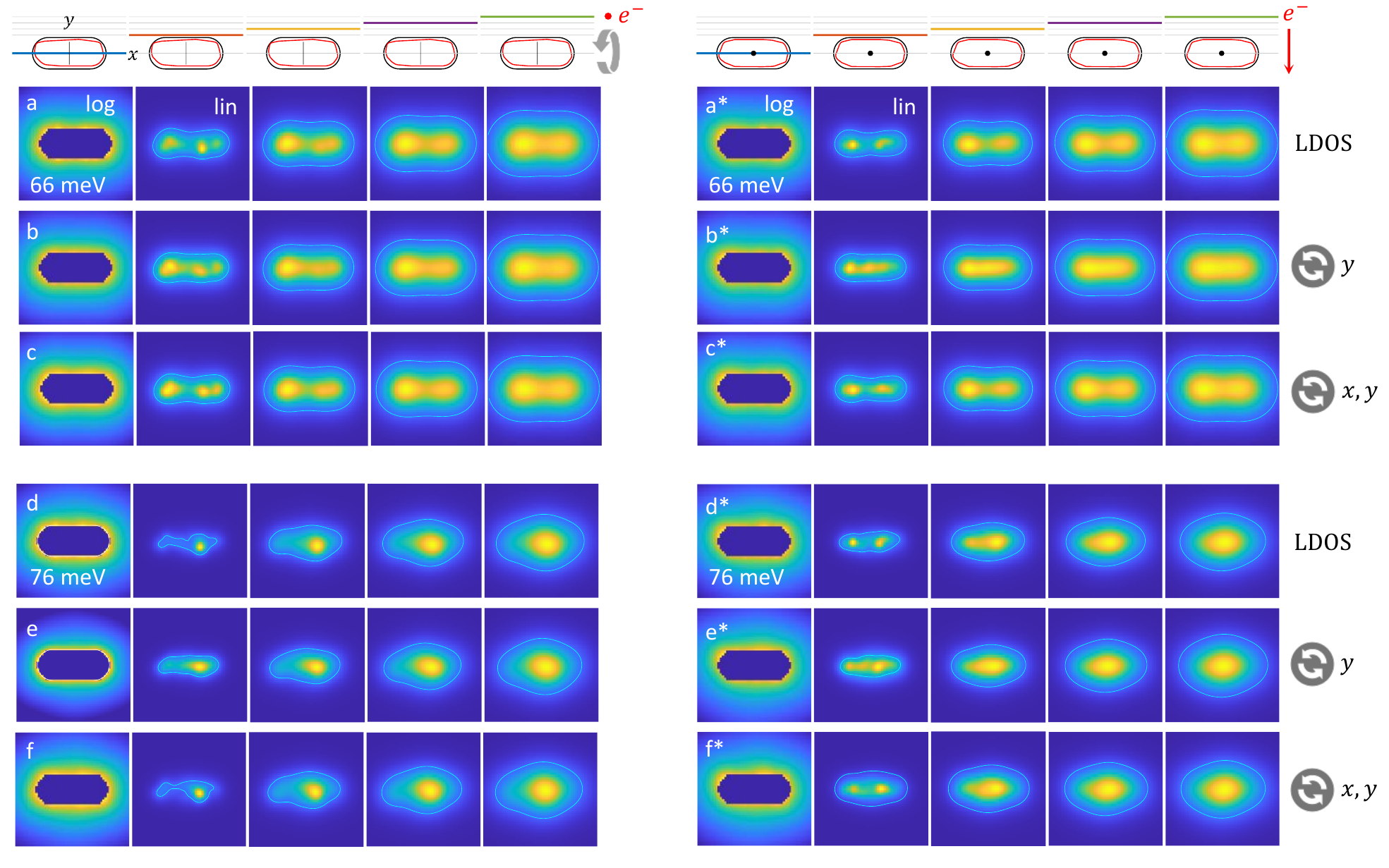}}
\caption{Same as Fig.~\ref{fig:ldos1} but for a rough nanorod and for the (a--c) dipolar and (d--f) quadrupolar rod resonance.  In the recon\-struction we consider $n=200$ reference modes for a smooth nanord (black contour shown on top) and $m=20$ modes to be reconstructed.  We compare \textsc{ldos} values in planes (a--f) parallel and (a*--f*) perpendicular to the electron propagation direction.  The lines and dots in the rod center indicate the tilt axis.  In panels (b,e) we conisder a tilt series where the nanoparticle is rotated around the $y$-axis only, whereas in panels (c,f) we additionally consider a rotation around the $x$-axis by 90$^\circ$ followed by the same tilt series around $y$.}\label{fig:ldos3}
\end{figure*}

Figures~\ref{fig:ldos1}(b,c) show results for the reconstructed \textsc{ldos} using (b) $n=100$ and (c) $n=20$ Laplace-Beltrami reference modes.  Further away from the rod, the simulated and reconstructed results agree well for both truncation numbers $n$.  For distances closer to the rod, the larger number of eigenmodes provides better agreement.  This is in accordance to our previous reasoning that oscillating dipoles close to the rod couple to a larger number of eigenmodes, and thus a larger number of modes is needed for the reconstruction.

In Fig.~\ref{fig:ldos2} we give a quantitative comparison between the simulated (full lines) and reconstructed (dashed lines) \textsc{ldos} values for cuts along the long rod axis and for dipole positions outside the nanoparticle.  The true \textsc{ldos} enhancement would depend on the actual size of the nanorod, for simplicity we give the results in arbitrary units.  Also the reconstruced \textsc{ldos} cuts are scaled by a constant factor, where it is not obvious how this factor could be obtained in absence of \textsc{eels} loss probabilities given in absolute numbers.  We here do not enter into the question of how to extract the absolute numbers of the reconstructed \textsc{ldos}.  Besides this unknown prefactor, the simulated and reconstructed \textsc{ldos} values agree extremely well, with the possible exception of the smallest distances where a larger number of eigenmodes might be needed.

Finally, in the remaining panels of Fig.~\ref{fig:ldos1} we compare the simulated and reconstructed \textsc{ldos} for the (d,e) quadrupolar rod mode and the (f,g) multitude of modes.  It can be seen that the reconstruction works well for the quadrupolar mode.  Comparison with results for $n=20$ (not shown) reveal that in this case a larger number of eigenmodes is strictly needed to obtain good agreement.  For the multitude of modes shown in panels (f,g) the agreement between simulation and reconstruction is reasonable but not overly good.  In particular for the smallest distances the reconstructed maps show sharp or asymmetric features, which are absent in the simulated maps.  From these results we conclude that the \textsc{ldos} reconstruction works best for loss peaks that are governed by a few modes only.  

\begin{figure}[t]
\centerline{\includegraphics[width=\columnwidth]{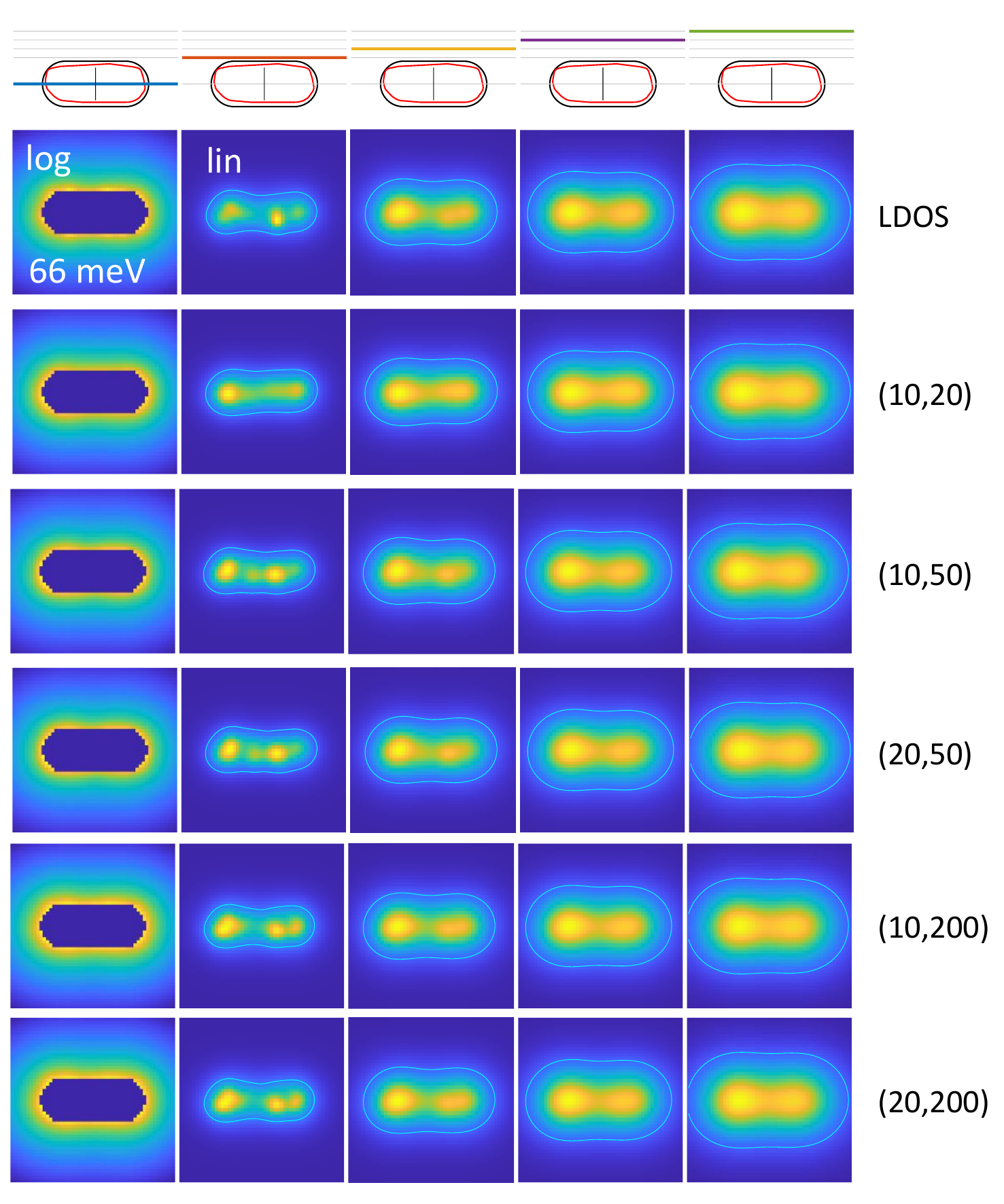}}
\caption{\textsc{ldos} maps for dipole mode of rough nanorod and for different $(m,n)$ cutoffs used in the optimization.  Here $m$ is the number of eigenpotentials to be reconstructed and $n$ is the number of basis modes.  As can be seen, the reconstruced \textsc{ldos} does not depend decisevly on the chosen parameters.}\label{fig:ldos4}
\end{figure}

\subsection{Rough rod}

The case of the rough rod shown in Fig.~\ref{fig:spectra} is considerably more difficult.  We keep considering the \textit{same} reference modes as for the smooth rod, and select the reference boundary $\partial\Omega_0$ such that it fully encapsulated the boundary $\partial\Omega$ of the rough rod.  Note that this reference boundary is identical to the one of the smooth rod.  Fig.~\ref{fig:cost} shows for the dipolar mode the simulated (``experimental'') \textsc{eels} maps and the reprojected ones.  The relative error between these maps is small throughout.

In Fig.~\ref{fig:ldos3} we show the simulated and reconstructed \textsc{ldos} maps for the rough nanorod.  We compare different planes that are (a--f) parallel and (a*--f*) perpendicular to the electron beam direction.  The main difference between these two configurations is that in the parallel case we reconstruct the \textsc{ldos} throughout in regions through which swift electrons have traveled.  In contrast, for the perpendicular case we reconstruct the \textsc{ldos} also in planes above the nanoparticle through which no electron has traveled because of our restriction to aloof trajectories.

Let us consider the parallel case first.  With the possible exception of the smallest distance, the agreement between simulated and reconstructed \textsc{ldos} maps is extremely good, both for the dipolar and quadrupolar mode.  Both asymmetries as well as hot spots, caused by localized fields in the vicinity of protrusions of the rough rod, are well reproduced by our tomography scheme.  Things somewhat change for the perpendicular geometry shown in panels (b*) and (e*), where the comparison is reasonable but not overly good.  We performed additional simulations where the tilt series for $\Gamma_{\rm exp}$ is complemented by \textsc{eels} maps where the nanorod is first rotated around the $x$-axis by $90^\circ$ before being submitted to the same tilt series around $y$.  As can be seen in panels (c*) and (f*), with this procedure we again obtain extremely good agreement between simulated and reconstructed \textsc{ldos} maps.  This shows that our tomography scheme works best for regions through which electrons have traveled.  

We finally investigate in Fig.~\ref{fig:ldos4} the impact of the cutoff parameters $(m,n)$ on the reconstructed \textsc{ldos} for the dipole mode.  Recall that $m$ is the number of eigenpotentials to be reconstructed, and $n$ is the cutoff parameter for the basis functions.  Close to the particle (second column) a larger number $n$ of basis states leads to a better agreement with the simulated \textsc{ldos}, shown in the first row.  When moving away from the nanoparticle, the agreement between simulated and reconstruced \textsc{ldos} maps is very good for all chosen simulation parameters.  This demonstrates that our reconstruction scheme is robust and that the optimization results do not depend decisively on the input parameters.

\section{Discussion}\label{sec:discussion}

In the previous sections we have presented the methodology of our tomographic reconstruction scheme and have investigated the approach for prototypical nanophotonic structures.  In this section we start by discussing our scheme within a broader context, and then address limitations, dos and dont's, as well as extensions of our tomographic reconstructions.

\subsection{Working principle}

\begin{figure}[t]
\centerline{\includegraphics[width=\columnwidth]{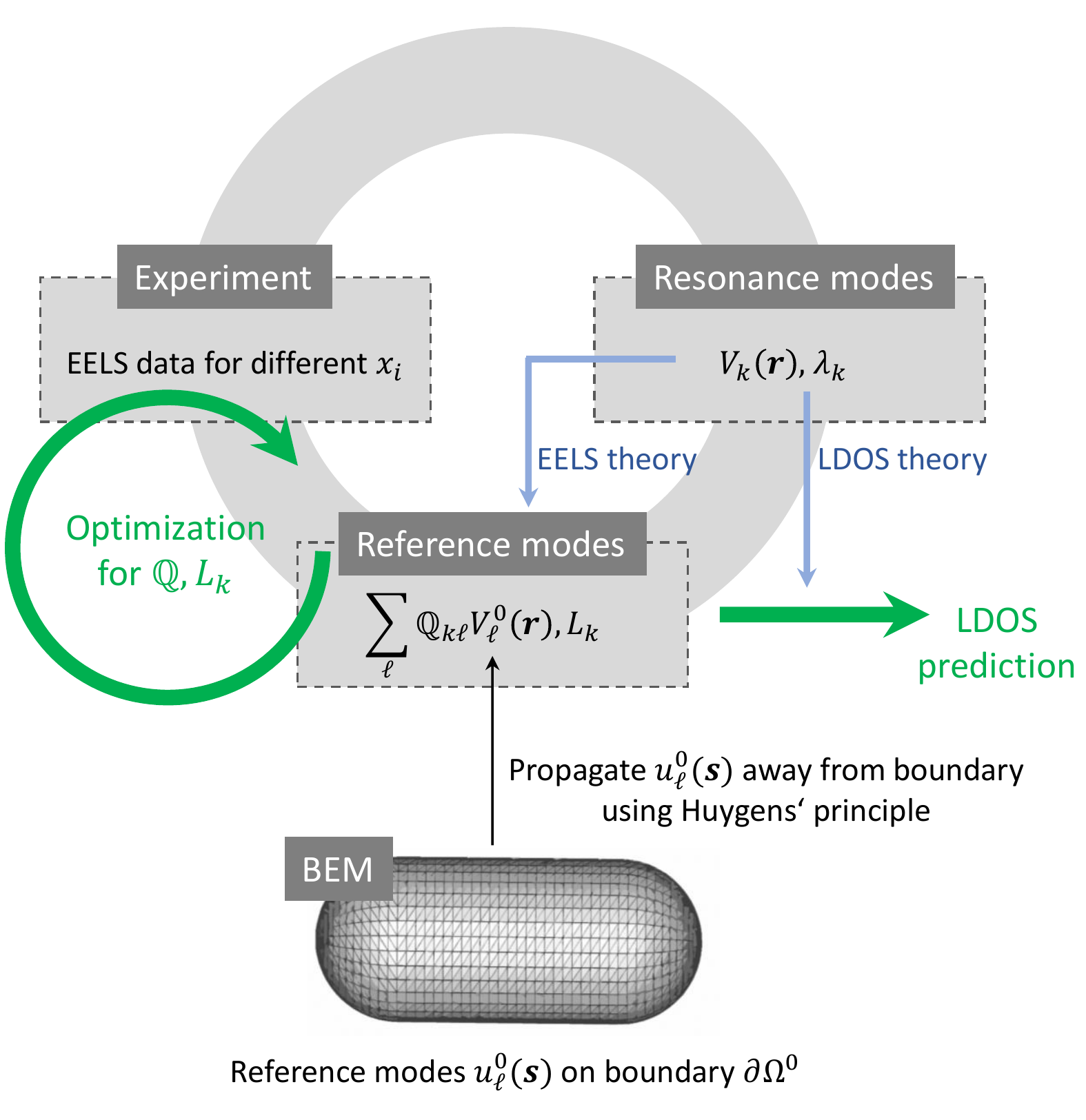}}
\caption{Working principle of our tomography scheme.  The approach consists of the triad of experiment, resonance modes, and references modes.  The experimental \textsc{eels} maps for various tilt angles provide the basic resource for the reconstruction of the photonic environment.  The resonance modes are used to formulate the theory underlying the reconstruction, the reference modes provide the parametrization of the photonic environment and are used for the actual reconstruction.  The parameters $L_k$, $\mathbb Q$ are obtained through an optimization procedure in order to minimize the difference between the measured and reprojected maps.  As the potentials outside the nanoparticle are solutions of Laplace's equation, we can employ a boundary element method (\textsc{bem}) scheme to express the potentials through their values on a boundary.}\label{fig:basic}
\end{figure}

The basic working principle of our tomographic reconstruction is shown in Fig.~\ref{fig:basic} and consists of the triad formed by experiment, resonance modes, and reference modes.  In short, the resonance modes are needed to formulate the theory, and the reference modes to provide a parametrization of the nanophotonic environment and to perform the actual reconstruction.  The experimental data are the primary resource for the reconstruction.  For this reason, the quality of the experimental data directly influences the quality of the tomographic reconstruction.  Some further considerations about experiments will be given below.

\subsubsection{Resonance modes}

The nanophotonic environment outside the nanoparticle is fully characterized in terms of the the reflected Green's function of Eq.~\eqref{eq:grefl1}, which we repeat here in compact form
\begin{equation}\label{eq:grefl2}
  G_{\rm refl}(\bm r,\bm r')=-\sum_k V_k(\bm r)\big(M_k+iL_k\big)V_k(\bm r')\,.
\end{equation}
$M_k$ and $L_k$ are the real and imaginary parts of the term given in brackets of Eq.~\eqref{eq:grefl1}.  The eigenpotentials $V_k(\bm r)$ provide the preferred physical basis, only with this basis the reflected Green's function can be written in the diagonal form of Eq.~\eqref{eq:grefl2}.  A similar decomposition of the Green's function can be also obtained in the retarded case when using quasinormal modes~\cite{leung:94,lalanne:19,kristensen:20}, as will be discussed below.  For this reason, from here on we use the more general expression of resonance modes rather than geometric eigenmodes, for which we have developed our theory so far. 

With these modes, both the \textsc{eels} loss probability of Eq.~\eqref{eq:eels2} as well as the power dissipation of an oscillating dipole, Eq.~\eqref{eq:ldos2}, can be written as the sum over individual loss channels.  With any other basis one would obtain some kind of mixing between different modes.  This particular form has the additional advantage that the lineshape function $L_k$ is always positive, at least for lossy materials, which can be used in our optimization procedure as a constraint, see Appendix~\ref{sec:ortho}.  Note that our tomography scheme only allows for the reconstruction of $L_k$, which accounts for the loss properties of the nanophotonic environment, but not for the propagation properties described by $M_k$.  As \textsc{eels} and \textsc{ldos} account for energy losses of electrons and oscillating dipoles, respectively, this is not a problem here.  However, additional experimental input or a reconstruction for various loss energies together with a Kramer-Kronig analysis would be needed for a reconstruction of $M_k$. 

To summarize this part, resonance modes are needed to formulate the abstract theory, without making contact to the actual shape or composition of the nanoparticle.  Without resonance modes it would be unclear which properties of the nanophotonic environment govern \textsc{eels} and \textsc{ldos}, and which properties can be reconstructed using an inverse scheme.  However, at no point of our approach we require explicit knowledge of the actual form of the resonance modes or lineshape functions.

\subsubsection{Reference modes}

The reference modes are the device needed for the actual reconstruction.  They allow for a suitable paramet\-rization of the nanophotonic environment, where the viable parameters can be extracted from the optimization loop using the experimental and reprojected \textsc{eels} maps.  In principle, for a complete basis the choice of the reference modes is irrelevant.  However, in all practical cases one has to truncate the basis, which should thus be based on an educated guess and should include the gross features of the expected resonance modes from the outset.

\subsubsection{Boundary element method approach}

Although all our calculations presented here and elsewhere~\cite{hoerl.prl:13,hoerl:15,hoerl:17,haberfehlner:17,li:21} have been performed using a boundary element method (\textsc{bem}) approach, it doesn't play an exceptional role in our tomography scheme.  The reference modes are fixed by specifying their values on a properly chosen reference boundary, see Fig.~\ref{fig:basic}.  Away from the boundary the modes propagate according to Laplace's equation, see Eq.~\eqref{eq:eigpot}, or the source-free Maxwell's equations in the retarded case.  This propagation is reminiscent of Huygens' principle for the wavefront propagation in free space and can be well described within \textsc{bem}, but otherwise our tomography makes no particular use of it.

\subsection{Frequently asked questions}

All our cards are on the table now.  Up to here, we have presented and examined our tomography approach in some detail, and have put it into a broader context.  However, a number of open or not fully clear issues remains.  In the following we address these issues in the form of frequently asked questions.  As will become apparent, only some of these questions can be answered definitely while others remain open.  In this sense, the following discussion is meant to summarize our present understanding of the field, to make aware where things can go wrong, and to identify directions for future research. 

\textit{How much preknowledge is needed?}  Any tomography or inverse scheme requires some sort of preknowledge, less preknowledge usually makes an approach more general and powerful.  In our case, we assume that the nanophotonic enviromnent can be expressed in terms of resonance modes, and that the potentials away from the boundary propagate as solutions of Laplace's equation.  

\textit{How many reference modes (and which) are required?}  For any practical reconstruction one has to truncate the basis.  The proper choice and truncation of the reference basis thus enters as an additional preknowledge.  For spectrally isolated resonances often a few tens of modes suffice, while in other cases up to hundred modes might be needed.  We are not aware of any general approach for determining the correct number of reference modes, so we advice potential users to vary the number and to obtain the best cutoff parameter on a case-to-case basis.

\textit{Are more reference modes always better?}  More modes slow down the optimization and require more iterations until reaching convergence.  With the simulated \textsc{eels} data used in this paper the quality of the reconstruction did not depend decisively on the truncation number.  However, things might change for noisy experimental data because higher modes come along with spatially more localized potential variations, and at some point one might end up reconstructing these noisy features rather than the real physical ones.

\textit{What is the typical computational cost?}  Depending on the truncation number, typical optimization times range from one to several minutes on a normal computer.  The code developments on top of a \textsc{bem} solver, such as the \textsc{nanobem} one~\cite{hohenester.cpc:22}, are moderate.  In the future we consider publishing our code to make it accessible to interested users.

\textit{Are there differences between \textsc{eels} and \textsc{ldos}?}  The question is somewhat odd, obviously \textsc{eels} accounts for the energy loss of swift electrons and \textsc{ldos} accounts for the enhancement of the decay rate of oscillating dipoles.  On the other hand, the interaction potential of the swift electron to the nanoparticle has a $\log(r)$ spatial dependence while the dipole has a $\nicefrac 1{r^2}$ dependence.  For this reason, \textsc{eels} maps are governed by the long-range features of the potential and \textsc{ldos} maps by the short-range features.  The prediction of \textsc{ldos} maps from experimental \textsc{eels} maps is thus a challenging and difficult task, and the good agreement between simulated and reprojected \textsc{ldos} maps reported in this work should not be taken as granted.

\textit{Does the optimization always succeed?}  In all reconstructions considered in this work the optimization algorithm ended up in a minimum.  However, there is no guarantee that this is a global minimum.  Our results never depended decisively on the initial values for $L_k$, $\mathbb{Q}$, which we initially set equal to one.  Note that zeros would be a bad choice because the derivatives of the cost function with respect to the optimization parameters would equate to zero then.  We generally recommend using quasi-Newton optimizations rather than conjugate gradient ones, because they typically access larger portions of the parameter space.

\textit{How much experimental input is needed?}  We will not give too much advice on the experiments here, interested readers might consult our previous work~\cite{hoerl:17,haberfehlner:17,li:21} to see what worked for us.  Depending on the electron microscope, contamination might play a role and might limit the amount of experimental data.  As has been discussed before, our tomography seems to work best for regions through which swift electrons have traveled.  The reconstruction of blind spots is possible, but the results should be handled with care.

\textit{How to address retardation?}  For surface phonon polaritons the quasistatic approximation works perfectly, but things might be more problematic for the reconstruction of other surface modes, such as surface plasmon polaritons.  In the past we have developed a methodology for surface plasmon tomography including retardation, and have applied the scheme to experimental \textsc{eels} data~\cite{hoerl:15,hoerl:17,haberfehlner:17}.  There we used a bi-orthogonal basis, which shares many features with the resonance modes presented here, but provides no justification for a strictly positive lineshape function.  For this reason, we opted for a compressed sensing optimization that favors expansions with as few modes as possible, where luckily all of them contributed with a positive weight to the loss probability.  In light of our present analysis, we suggest a slight modification of our previous approach.  First, in the retarded case the preferred basis is given by quasinormal modes~\cite{leung:94,lalanne:19,kristensen:20,hohenester.cpc:22}, which have received considerable interest recently.  With these modes we can decompose the dyadic Green's tensor for the full Maxwell's equations in a form similar to Eq.~\eqref{eq:grefl2}, and a tomographic reconstruction should be possible along the lines sketched in the present work.  There remain a number of open issues, such as the proper choice of the reference modes or the consideration of complex mode functions, but we don't foresee any major roadblock.  As a side remark, it is no surprise that the decomposition of the reflected Green's function in terms of resonance modes looks similar in the quasistatic and retarded case: such decompositions are in the spirit of generic singular value decompositions, where the special structure is due to the symmetry property of Green's function originating from the reciprocity theorem of optics.

\textit{How to address membranes and grids?}  Membranes or grids are needed in experiment to support the nanoparticle.  One might wonder about the consequences of such a support in our tomographic reconstruction.  First, modifications of the resonance energies or surface charge distributions of the surface phonon polaritons can be already properly accounted for with the present approach, as has been demonstrated in~\cite{li:21}.  However, in principle also the free-space Green's function of Eq.~\eqref{eq:green2} should be modified to account for the dielectric environment in presence of a support.  This modified Green's function should be used in Eq.~\eqref{eq:eigpot} to propagate away the potentials from the boundary.  It is the resulting modification of the electron-nanoparticle interaction that has to be considered.  Whether this modification has noticeable influence on the results has to be seen.

To summarize, \textsc{eels} tomography has become a successful scheme for reconstructing the three-dimensional photonic environment of nanoparticles with high spatial and energy resolution.  In the past several case studies for plasmonic and photonic nanostructures have provided beautiful results, which would have been hard to achieve with other techniques.  Yet, we feel that there is still enough room for improvements and further investigations.  In this paper we have given an in-depth study of a prototypical nanophotonic system and have demonstrated that tomographic reconstructions work reliably and without major difficulties, at least for systems where the quasistatic approximation can be employed.  We hope that this will motivate more research groups to enter the field, to investigate their systems with the tools presented here, and to continue developing \textsc{eels} tomography with further improvements.

\section*{Acknowledgements}

This project has received funding from the European Union’s Horizon 2020 Research and Innovation Program under
grant agreement 823717 (ESTEEM3), from the Austrian Science Fund FWF under project P 31264 and by NAWI Graz.

\begin{appendix}

\section{}\label{sec:ortho}

In this appendix we first derive Eq.~\eqref{eq:eigpotref1}.  We denote the nanoparticle boundary with $\partial\Omega$ and the geometric eigenmodes with $u_k(\bm s)$.  These eigenmodes form a complete set of basis functions and fulfill the orthogonality relation~\cite{ouyang:89,boudarham:12,hohenester:20}
\begin{equation}\label{eq:ortho}
  \oint_{\partial\Omega}\frac{u_k(\bm s)u_{k'}(\bm s')}{4\pi|\bm s-\bm s'|}\,dSdS'=\delta_{kk'}\,.
\end{equation}
Let $u^0_\ell(\bm s_0)$ be the reference basis functions, which are assumed to fulfill a similar orthogonality relation, 
\begin{equation}\label{eq:orthoref}
  \oint_{\partial\Omega}\frac{u^0_\ell(\bm s)u^0_{\ell'}(\bm s')}{4\pi|\bm s-\bm s'|}\,dS dS'=
  \delta_{\ell\ell'}\,.
\end{equation}
This can be always achieved for a set of basis functions using a Gram-Schmidt-type orthogonalization.  As $u^0_\ell(\bm s)$ form a complete basis, we can expand the eigenmodes via
\begin{equation}
  u_k(\bm s)=\sum_\ell\mathbb{Q}_{k\ell}u_\ell^0(\bm s)\,.
\end{equation}
Inserting this expression into Eq.~\eqref{eq:ortho} and using the orthogonality relation of Eq.~\eqref{eq:orthoref}, we then immediately observe that $\mathbb{Q}$ is an orthogonal matrix.  In our computational approach we employ Cayley's parameterization for orthogonal matrices
\begin{equation}\label{eq:cayley}
  \mathbb{Q}=\big(\openone+\mathbb{X}\big)\big(\openone-\mathbb{X}\big)^{-1}\,,
\end{equation}
where $\mathbb{X}$ is a skew-symmetric matrix with $x_{ij}=-x_{ji}$.  Eq.~\eqref{eq:cayley} has the advantage that one can perform the derivative $\nicefrac{\partial\mathbb{Q}}{\partial x_{ij}}$ analytically.  To show this, we start from
\begin{equation}\label{eq:cayley2}
  \frac{\partial\mathbb{Q}}{\partial x}=
  \frac{\partial\mathbb{X}}{\partial x}\big(\openone-\mathbb{X}\big)^{-1}+
  \big(\openone+\mathbb{X}\big)\frac\partial{\partial x}\big(\openone-\mathbb{X}\big)^{-1}\,,
\end{equation}
where for notational clarity we have suppressed the subscripts of $x$.  To evaluate the second term on the right hand side, we differentiate $(\openone-\mathbb{X})(\openone-\mathbb{X})^{-1}=\openone$  with respect to $x$.  After some manipulations this leads to
\begin{displaymath}
  \frac\partial{\partial x}\big(\openone-\mathbb{X}\big)^{-1}=
  \big(\openone-\mathbb{X}\big)^{-1}\frac{\partial\mathbb{X}}{\partial x}
  \big(\openone-\mathbb{X}\big)^{-1}\,.
\end{displaymath}
Insertion into Eq.~\eqref{eq:cayley2} gives
\begin{displaymath}
  \frac{\partial\mathbb{Q}}{\partial x}=
  \frac{\partial\mathbb{X}}{\partial x}\big(\openone-\mathbb{X}\big)^{-1}+
  \big(\openone+\mathbb{X}\big)\big(\openone-\mathbb{X}\big)^{-1}\frac{\partial\mathbb{X}}{\partial x}\big(\openone-\mathbb{X}\big)^{-1}\,.
\end{displaymath}
We next use $(\openone+\mathbb{X})(\openone-\mathbb{X})^{-1}=(\openone-\mathbb{X})^{-1}(\openone+\mathbb{X})$, which can be easily proven using $\openone+\mathbb{X}=2\openone-(\openone-\mathbb{X})$ and that both terms on the right hand side commute with $(\openone-\mathbb{X})^{-1}$.  We then arrive at our final expression
\begin{equation}\label{eq:cayleyderiv}
  \frac{\partial Q}{\partial x}=2\big(\openone-\mathbb{X}\big)^{-1}
  \frac{\partial X}{\partial x}\big(\openone-\mathbb{X}\big)^{-1}\,.
\end{equation}
In our optimization of the cost function we express the lineshape function through $L_k=s_k^2$, which guarantees that $L_k$ is always positive.  The optimization algorithm can be significantly accelerated by providing in addition to the value of the cost function also the derivatives with respect to the optimization parameters.  Using Eq.~\eqref{eq:eels3} together with Eq.~\eqref{eq:cayleyderiv}, the derivative of the cost function \eqref{eq:cost} with respect to the parameters $s_k$ and $x_{k\ell}$ of the skew-symmetric matrix can be obtained analytically.  Things are considerably easier for a non-orthogonal matrix where the derivatives with respect to the matrix elements can be performed straightforwardly. 

\end{appendix}


\begin{thebibliography}{32}
\expandafter\ifx\csname natexlab\endcsname\relax\def\natexlab#1{#1}\fi
\expandafter\ifx\csname bibnamefont\endcsname\relax
  \def\bibnamefont#1{#1}\fi
\expandafter\ifx\csname bibfnamefont\endcsname\relax
  \def\bibfnamefont#1{#1}\fi
\expandafter\ifx\csname citenamefont\endcsname\relax
  \def\citenamefont#1{#1}\fi
\expandafter\ifx\csname url\endcsname\relax
  \def\url#1{\texttt{#1}}\fi
\expandafter\ifx\csname urlprefix\endcsname\relax\def\urlprefix{URL }\fi
\providecommand{\bibinfo}[2]{#2}
\providecommand{\eprint}[2][]{\url{#2}}

\bibitem[{\citenamefont{Novotny and Hecht}(2006)}]{novotny:06}
\bibinfo{author}{\bibfnamefont{L.}~\bibnamefont{Novotny}} \bibnamefont{and}
  \bibinfo{author}{\bibfnamefont{B.}~\bibnamefont{Hecht}},
  \emph{\bibinfo{title}{Principles of Nano-Optics}}
  (\bibinfo{publisher}{Cambridge University Press, Cambridge},
  \bibinfo{year}{2006}).

\bibitem[{\citenamefont{Hohenester}(2020)}]{hohenester:20}
\bibinfo{author}{\bibfnamefont{U.}~\bibnamefont{Hohenester}},
  \emph{\bibinfo{title}{Nano and Quantum Optics}}
  (\bibinfo{publisher}{Springer}, \bibinfo{address}{Cham, Switzerland},
  \bibinfo{year}{2020}).

\bibitem[{\citenamefont{Maier}(2007)}]{maier:07}
\bibinfo{author}{\bibfnamefont{S.~A.} \bibnamefont{Maier}},
  \emph{\bibinfo{title}{Plasmonics: Fundamentals and Applications}}
  (\bibinfo{publisher}{Springer}, \bibinfo{address}{Berlin},
  \bibinfo{year}{2007}).

\bibitem[{\citenamefont{Kliewer and Fuchs}(1974)}]{kliewer:74}
\bibinfo{author}{\bibfnamefont{K.}~\bibnamefont{Kliewer}} \bibnamefont{and}
  \bibinfo{author}{\bibfnamefont{R.}~\bibnamefont{Fuchs}},
  \emph{\bibinfo{title}{Theory of Dynamical Properties of Dielectric
  Surfaces}}, vol.~\bibinfo{volume}{27} (\bibinfo{publisher}{Wiley},
  \bibinfo{address}{Weinheim}, \bibinfo{year}{1974}).

\bibitem[{\citenamefont{Caldwell et~al.}(2015)\citenamefont{Caldwell, Lindsay,
  Giannini, Vurgaftman, Reinecke, Maier, and Glembocki}}]{caldwell:15}
\bibinfo{author}{\bibfnamefont{J.~D.} \bibnamefont{Caldwell}},
  \bibinfo{author}{\bibfnamefont{L.}~\bibnamefont{Lindsay}},
  \bibinfo{author}{\bibfnamefont{V.}~\bibnamefont{Giannini}},
  \bibinfo{author}{\bibfnamefont{I.}~\bibnamefont{Vurgaftman}},
  \bibinfo{author}{\bibfnamefont{T.~L.} \bibnamefont{Reinecke}},
  \bibinfo{author}{\bibfnamefont{S.~A.} \bibnamefont{Maier}}, \bibnamefont{and}
  \bibinfo{author}{\bibfnamefont{O.~J.} \bibnamefont{Glembocki}},
  \bibinfo{journal}{Nanophotonics} \textbf{\bibinfo{volume}{4}},
  \bibinfo{pages}{44} (\bibinfo{year}{2015}).

\bibitem[{\citenamefont{Barbillon}(2019)}]{barbillon:19}
\bibinfo{author}{\bibfnamefont{H.}~\bibnamefont{Barbillon}},
  \bibinfo{journal}{Materials} \textbf{\bibinfo{volume}{12}},
  \bibinfo{pages}{1502} (\bibinfo{year}{2019}).

\bibitem[{\citenamefont{Nelayah et~al.}(2007)\citenamefont{Nelayah, Kociak,
  Stephan, Garc{\'i}{a de Abajo}, Tence, Henrard, Taverna, Pastoriz{a-Santos},
  Li{z-Martin}, and Colliex}}]{nelayah:07}
\bibinfo{author}{\bibfnamefont{J.}~\bibnamefont{Nelayah}},
  \bibinfo{author}{\bibfnamefont{M.}~\bibnamefont{Kociak}},
  \bibinfo{author}{\bibfnamefont{O.}~\bibnamefont{Stephan}},
  \bibinfo{author}{\bibfnamefont{F.~J.} \bibnamefont{Garc{\'i}{a de Abajo}}},
  \bibinfo{author}{\bibfnamefont{M.}~\bibnamefont{Tence}},
  \bibinfo{author}{\bibfnamefont{L.}~\bibnamefont{Henrard}},
  \bibinfo{author}{\bibfnamefont{D.}~\bibnamefont{Taverna}},
  \bibinfo{author}{\bibfnamefont{I.}~\bibnamefont{Pastoriz{a-Santos}}},
  \bibinfo{author}{\bibfnamefont{L.~M.} \bibnamefont{Li{z-Martin}}},
  \bibnamefont{and} \bibinfo{author}{\bibfnamefont{C.}~\bibnamefont{Colliex}},
  \bibinfo{journal}{Nature Phys.} \textbf{\bibinfo{volume}{3}},
  \bibinfo{pages}{348} (\bibinfo{year}{2007}).

\bibitem[{\citenamefont{Kociak and Stephan}(2014)}]{kociak:14}
\bibinfo{author}{\bibfnamefont{M.}~\bibnamefont{Kociak}} \bibnamefont{and}
  \bibinfo{author}{\bibfnamefont{O.}~\bibnamefont{Stephan}},
  \bibinfo{journal}{Chem. Soc. Rev.} \textbf{\bibinfo{volume}{53}},
  \bibinfo{pages}{3865} (\bibinfo{year}{2014}).

\bibitem[{\citenamefont{Colliex et~al.}(2016)\citenamefont{Colliex, Kociak, and
  Stephan}}]{colliex:16}
\bibinfo{author}{\bibfnamefont{C.}~\bibnamefont{Colliex}},
  \bibinfo{author}{\bibfnamefont{M.}~\bibnamefont{Kociak}}, \bibnamefont{and}
  \bibinfo{author}{\bibfnamefont{O.}~\bibnamefont{Stephan}},
  \bibinfo{journal}{Ultramicroscopy} \textbf{\bibinfo{volume}{162}},
  \bibinfo{pages}{A1} (\bibinfo{year}{2016}).

\bibitem[{\citenamefont{Polman et~al.}(2019)\citenamefont{Polman, Kociak, and
  Garc{ia de Ab}ajo}}]{polman:19}
\bibinfo{author}{\bibfnamefont{A.}~\bibnamefont{Polman}},
  \bibinfo{author}{\bibfnamefont{M.}~\bibnamefont{Kociak}}, \bibnamefont{and}
  \bibinfo{author}{\bibfnamefont{J.~F.} \bibnamefont{Garc{ia de Ab}ajo}},
  \bibinfo{journal}{Nature Materials} \textbf{\bibinfo{volume}{18}},
  \bibinfo{pages}{1158} (\bibinfo{year}{2019}).

\bibitem[{\citenamefont{Garc{\'i}{a de Abajo}}(2010)}]{garcia:10}
\bibinfo{author}{\bibfnamefont{F.~J.} \bibnamefont{Garc{\'i}{a de Abajo}}},
  \bibinfo{journal}{Rev. Mod. Phys.} \textbf{\bibinfo{volume}{82}},
  \bibinfo{pages}{209} (\bibinfo{year}{2010}).

\bibitem[{\citenamefont{Midgley and Duni{n-B}orkowski}(2009)}]{midgley:09}
\bibinfo{author}{\bibfnamefont{P.~A.} \bibnamefont{Midgley}} \bibnamefont{and}
  \bibinfo{author}{\bibfnamefont{R.~E.} \bibnamefont{Duni{n-B}orkowski}},
  \bibinfo{journal}{Nat. Mater.} \textbf{\bibinfo{volume}{8}},
  \bibinfo{pages}{271} (\bibinfo{year}{2009}).

\bibitem[{\citenamefont{H{\"o}rl et~al.}(2013)\citenamefont{H{\"o}rl,
  Tr{\"u}gler, and Hohenester}}]{hoerl.prl:13}
\bibinfo{author}{\bibfnamefont{A.}~\bibnamefont{H{\"o}rl}},
  \bibinfo{author}{\bibfnamefont{A.}~\bibnamefont{Tr{\"u}gler}},
  \bibnamefont{and}
  \bibinfo{author}{\bibfnamefont{U.}~\bibnamefont{Hohenester}},
  \bibinfo{journal}{Phys. Rev. Lett.} \textbf{\bibinfo{volume}{111}},
  \bibinfo{pages}{086801} (\bibinfo{year}{2013}).

\bibitem[{\citenamefont{Nicoletti et~al.}(2013)\citenamefont{Nicoletti, d{e la
  P}ena, Leary, Holland, Ducati, and Midgley}}]{nicoletti:13}
\bibinfo{author}{\bibfnamefont{O.}~\bibnamefont{Nicoletti}},
  \bibinfo{author}{\bibfnamefont{F.}~\bibnamefont{d{e la P}ena}},
  \bibinfo{author}{\bibfnamefont{R.~W.} \bibnamefont{Leary}},
  \bibinfo{author}{\bibfnamefont{D.~J.} \bibnamefont{Holland}},
  \bibinfo{author}{\bibfnamefont{C.}~\bibnamefont{Ducati}}, \bibnamefont{and}
  \bibinfo{author}{\bibfnamefont{P.~A.} \bibnamefont{Midgley}},
  \bibinfo{journal}{Nature} \textbf{\bibinfo{volume}{502}}, \bibinfo{pages}{80}
  (\bibinfo{year}{2013}).

\bibitem[{\citenamefont{H{\"o}rl et~al.}(2015)\citenamefont{H{\"o}rl,
  Tr{\"u}gler, and Hohenester}}]{hoerl:15}
\bibinfo{author}{\bibfnamefont{A.}~\bibnamefont{H{\"o}rl}},
  \bibinfo{author}{\bibfnamefont{A.}~\bibnamefont{Tr{\"u}gler}},
  \bibnamefont{and}
  \bibinfo{author}{\bibfnamefont{U.}~\bibnamefont{Hohenester}},
  \bibinfo{journal}{ACS Photonics} \textbf{\bibinfo{volume}{2}},
  \bibinfo{pages}{1429} (\bibinfo{year}{2015}).

\bibitem[{\citenamefont{H{\"o}rl et~al.}(2017)\citenamefont{H{\"o}rl,
  Haberfehlner, Tr{\"u}glerl, Schmidt, Hohenester, and Kothleitner}}]{hoerl:17}
\bibinfo{author}{\bibfnamefont{A.}~\bibnamefont{H{\"o}rl}},
  \bibinfo{author}{\bibfnamefont{G.}~\bibnamefont{Haberfehlner}},
  \bibinfo{author}{\bibfnamefont{A.}~\bibnamefont{Tr{\"u}glerl}},
  \bibinfo{author}{\bibfnamefont{F.}~\bibnamefont{Schmidt}},
  \bibinfo{author}{\bibfnamefont{U.}~\bibnamefont{Hohenester}},
  \bibnamefont{and}
  \bibinfo{author}{\bibfnamefont{G.}~\bibnamefont{Kothleitner}},
  \bibinfo{journal}{Nature Commun.} \textbf{\bibinfo{volume}{8}},
  \bibinfo{pages}{37} (\bibinfo{year}{2017}).

\bibitem[{\citenamefont{Haberfehlner et~al.}(2017)\citenamefont{Haberfehlner,
  Schmidt, Schaffernak, H{\"o}rl, Tr{\"u}gler, Hohenau, Krenn, Hohenester, and
  Kothleitner}}]{haberfehlner:17}
\bibinfo{author}{\bibfnamefont{G.}~\bibnamefont{Haberfehlner}},
  \bibinfo{author}{\bibfnamefont{F.~P.} \bibnamefont{Schmidt}},
  \bibinfo{author}{\bibfnamefont{G.}~\bibnamefont{Schaffernak}},
  \bibinfo{author}{\bibfnamefont{A.}~\bibnamefont{H{\"o}rl}},
  \bibinfo{author}{\bibfnamefont{A.}~\bibnamefont{Tr{\"u}gler}},
  \bibinfo{author}{\bibfnamefont{A.}~\bibnamefont{Hohenau}},
  \bibinfo{author}{\bibfnamefont{J.~R.} \bibnamefont{Krenn}},
  \bibinfo{author}{\bibfnamefont{U.}~\bibnamefont{Hohenester}},
  \bibnamefont{and}
  \bibinfo{author}{\bibfnamefont{G.}~\bibnamefont{Kothleitner}},
  \bibinfo{journal}{Nano. Lett.} \textbf{\bibinfo{volume}{17}},
  \bibinfo{pages}{6773} (\bibinfo{year}{2017}).

\bibitem[{\citenamefont{Li et~al.}(2021)\citenamefont{Li, Haberfehlner,
  Hohenester, Stephan, Kothleitner, and Kociak}}]{li:21}
\bibinfo{author}{\bibfnamefont{X.}~\bibnamefont{Li}},
  \bibinfo{author}{\bibfnamefont{G.}~\bibnamefont{Haberfehlner}},
  \bibinfo{author}{\bibfnamefont{U.}~\bibnamefont{Hohenester}},
  \bibinfo{author}{\bibfnamefont{O.}~\bibnamefont{Stephan}},
  \bibinfo{author}{\bibfnamefont{G.}~\bibnamefont{Kothleitner}},
  \bibnamefont{and} \bibinfo{author}{\bibfnamefont{M.}~\bibnamefont{Kociak}},
  \bibinfo{journal}{Science} \textbf{\bibinfo{volume}{371}},
  \bibinfo{pages}{1364} (\bibinfo{year}{2021}).

\bibitem[{\citenamefont{Garc{\'i}{a de Abajo} and Kociak}(2008)}]{garcia:08}
\bibinfo{author}{\bibfnamefont{F.~J.} \bibnamefont{Garc{\'i}{a de Abajo}}}
  \bibnamefont{and} \bibinfo{author}{\bibfnamefont{M.}~\bibnamefont{Kociak}},
  \bibinfo{journal}{Phys. Rev. Lett.} \textbf{\bibinfo{volume}{100}},
  \bibinfo{pages}{106804} (\bibinfo{year}{2008}).

\bibitem[{\citenamefont{Schuller et~al.}(2010)\citenamefont{Schuller, Barnard,
  Cai, Jun, White, and Brongersma}}]{schuller:10}
\bibinfo{author}{\bibfnamefont{J.~A.} \bibnamefont{Schuller}},
  \bibinfo{author}{\bibfnamefont{E.~S.} \bibnamefont{Barnard}},
  \bibinfo{author}{\bibfnamefont{W.}~\bibnamefont{Cai}},
  \bibinfo{author}{\bibfnamefont{Y.~C.} \bibnamefont{Jun}},
  \bibinfo{author}{\bibfnamefont{J.~S.} \bibnamefont{White}}, \bibnamefont{and}
  \bibinfo{author}{\bibfnamefont{M.~L.} \bibnamefont{Brongersma}},
  \bibinfo{journal}{Nature Mat.} \textbf{\bibinfo{volume}{9}},
  \bibinfo{pages}{193} (\bibinfo{year}{2010}).

\bibitem[{\citenamefont{Jackson}(1999)}]{jackson:99}
\bibinfo{author}{\bibfnamefont{J.~D.} \bibnamefont{Jackson}},
  \emph{\bibinfo{title}{Classical Electrodynamics}}
  (\bibinfo{publisher}{Wiley}, \bibinfo{address}{New York},
  \bibinfo{year}{1999}).

\bibitem[{\citenamefont{Ouyang and Isaacson}(1989)}]{ouyang:89}
\bibinfo{author}{\bibfnamefont{F.}~\bibnamefont{Ouyang}} \bibnamefont{and}
  \bibinfo{author}{\bibfnamefont{M.}~\bibnamefont{Isaacson}},
  \bibinfo{journal}{Phil. Mag. B} \textbf{\bibinfo{volume}{60}},
  \bibinfo{pages}{481} (\bibinfo{year}{1989}).

\bibitem[{\citenamefont{Boudarham and Kociak}(2012)}]{boudarham:12}
\bibinfo{author}{\bibfnamefont{G.}~\bibnamefont{Boudarham}} \bibnamefont{and}
  \bibinfo{author}{\bibfnamefont{M.}~\bibnamefont{Kociak}},
  \bibinfo{journal}{Phys. Rev. B} \textbf{\bibinfo{volume}{85}},
  \bibinfo{pages}{245447} (\bibinfo{year}{2012}).

\bibitem[{\citenamefont{Press et~al.}(2002)\citenamefont{Press, Teukolsky,
  Vetterling, and Flannery}}]{press:02}
\bibinfo{author}{\bibfnamefont{W.~H.} \bibnamefont{Press}},
  \bibinfo{author}{\bibfnamefont{S.~A.} \bibnamefont{Teukolsky}},
  \bibinfo{author}{\bibfnamefont{W.~T.} \bibnamefont{Vetterling}},
  \bibnamefont{and} \bibinfo{author}{\bibfnamefont{B.~P.}
  \bibnamefont{Flannery}}, \emph{\bibinfo{title}{Numerical Recipes in {C+\!+}:
  The Art of Scientific Computing}} (\bibinfo{publisher}{Cambridge Univ.
  Press}, \bibinfo{address}{Cambridge}, \bibinfo{year}{2002}),
  \bibinfo{edition}{2nd} ed.

\bibitem[{\citenamefont{Louren{o-M}artins and Kociak}(2017)}]{lourenco:17}
\bibinfo{author}{\bibfnamefont{H.}~\bibnamefont{Louren{o-M}artins}}
  \bibnamefont{and} \bibinfo{author}{\bibfnamefont{M.}~\bibnamefont{Kociak}},
  \bibinfo{journal}{Phys. Rev. X} \textbf{\bibinfo{volume}{7}},
  \bibinfo{pages}{041059} (\bibinfo{year}{2017}).

\bibitem[{\citenamefont{Tr{\"u}gler et~al.}(2011)\citenamefont{Tr{\"u}gler,
  Tinguely, Krenn, Hohenau, and Hohenester}}]{truegler.prb:11}
\bibinfo{author}{\bibfnamefont{A.}~\bibnamefont{Tr{\"u}gler}},
  \bibinfo{author}{\bibfnamefont{J.~C.} \bibnamefont{Tinguely}},
  \bibinfo{author}{\bibfnamefont{J.~R.} \bibnamefont{Krenn}},
  \bibinfo{author}{\bibfnamefont{A.}~\bibnamefont{Hohenau}}, \bibnamefont{and}
  \bibinfo{author}{\bibfnamefont{U.}~\bibnamefont{Hohenester}},
  \bibinfo{journal}{Phys. Rev. B} \textbf{\bibinfo{volume}{83}},
  \bibinfo{pages}{081412(R)} (\bibinfo{year}{2011}).

\bibitem[{\citenamefont{Hohenester et~al.}(2022)\citenamefont{Hohenester,
  Reichelt, and Unger}}]{hohenester.cpc:22}
\bibinfo{author}{\bibfnamefont{U.}~\bibnamefont{Hohenester}},
  \bibinfo{author}{\bibfnamefont{N.}~\bibnamefont{Reichelt}}, \bibnamefont{and}
  \bibinfo{author}{\bibfnamefont{G.}~\bibnamefont{Unger}},
  \bibinfo{journal}{Comp. Phys. Comp.} \textbf{\bibinfo{volume}{276}},
  \bibinfo{pages}{108337} (\bibinfo{year}{2022}).

\bibitem[{\citenamefont{Hohenester et~al.}(2018)\citenamefont{Hohenester,
  A.~Tr\"ugler, Batson, and Lagos}}]{hohenester.prb:18}
\bibinfo{author}{\bibfnamefont{U.}~\bibnamefont{Hohenester}},
  \bibinfo{author}{\bibfnamefont{A.}~\bibnamefont{A.~Tr\"ugler}},
  \bibinfo{author}{\bibfnamefont{P.~E.} \bibnamefont{Batson}},
  \bibnamefont{and} \bibinfo{author}{\bibfnamefont{M.~J.} \bibnamefont{Lagos}},
  \bibinfo{journal}{Phys. Rev. B} \textbf{\bibinfo{volume}{97}},
  \bibinfo{pages}{165418} (\bibinfo{year}{2018}).

\bibitem[{\citenamefont{Bobenko and Springborn}(2007)}]{bobenko:07}
\bibinfo{author}{\bibfnamefont{A.~I.} \bibnamefont{Bobenko}} \bibnamefont{and}
  \bibinfo{author}{\bibfnamefont{B.~A.} \bibnamefont{Springborn}},
  \bibinfo{journal}{Discrete Comput. Geom.} \textbf{\bibinfo{volume}{38}},
  \bibinfo{pages}{740} (\bibinfo{year}{2007}).

\bibitem[{\citenamefont{Leung et~al.}(1994)\citenamefont{Leung, Liu, and
  Young}}]{leung:94}
\bibinfo{author}{\bibfnamefont{P.~T.} \bibnamefont{Leung}},
  \bibinfo{author}{\bibfnamefont{S.~Y.} \bibnamefont{Liu}}, \bibnamefont{and}
  \bibinfo{author}{\bibfnamefont{K.}~\bibnamefont{Young}},
  \bibinfo{journal}{Phys. Rev. A} \textbf{\bibinfo{volume}{49}},
  \bibinfo{pages}{3057} (\bibinfo{year}{1994}).

\bibitem[{\citenamefont{Lalanne et~al.}(2019)\citenamefont{Lalanne, Yan, Gras,
  Sauvan, Hugonin, Besbes, Demesy, Truong, Gralak, Zolla et~al.}}]{lalanne:19}
\bibinfo{author}{\bibfnamefont{P.}~\bibnamefont{Lalanne}},
  \bibinfo{author}{\bibfnamefont{W.}~\bibnamefont{Yan}},
  \bibinfo{author}{\bibfnamefont{A.}~\bibnamefont{Gras}},
  \bibinfo{author}{\bibfnamefont{C.}~\bibnamefont{Sauvan}},
  \bibinfo{author}{\bibfnamefont{J.-P.} \bibnamefont{Hugonin}},
  \bibinfo{author}{\bibfnamefont{M.}~\bibnamefont{Besbes}},
  \bibinfo{author}{\bibfnamefont{G.}~\bibnamefont{Demesy}},
  \bibinfo{author}{\bibfnamefont{M.~D.} \bibnamefont{Truong}},
  \bibinfo{author}{\bibfnamefont{B.}~\bibnamefont{Gralak}},
  \bibinfo{author}{\bibfnamefont{F.}~\bibnamefont{Zolla}},
  \bibnamefont{et~al.}, \bibinfo{journal}{J. Opt. Soc. Am. A}
  \textbf{\bibinfo{volume}{36}}, \bibinfo{pages}{686} (\bibinfo{year}{2019}).

\bibitem[{\citenamefont{Kristensen et~al.}(2020)\citenamefont{Kristensen,
  Herrmann, Intravaia, and Busch}}]{kristensen:20}
\bibinfo{author}{\bibfnamefont{P.~T.} \bibnamefont{Kristensen}},
  \bibinfo{author}{\bibfnamefont{K.}~\bibnamefont{Herrmann}},
  \bibinfo{author}{\bibfnamefont{F.}~\bibnamefont{Intravaia}},
  \bibnamefont{and} \bibinfo{author}{\bibfnamefont{K.}~\bibnamefont{Busch}},
  \bibinfo{journal}{Adv. Opt. Photon.} \textbf{\bibinfo{volume}{12}},
  \bibinfo{pages}{612} (\bibinfo{year}{2020}).

\end{thebibliography}

\end{document}